\pdfoutput=1

\documentclass[11pt,twoside,a4paper,cmspaper,final,collab]{cms-tdr}
\def\svnVersion{203614}\def\cmsCernNo{2013-084}\def\cmsCernDate{\today}\def\cmsMessage{Published in the Journal of High Energy Physics as \href{http://dx.doi.org/10.1007/JHEP07(2013)116}{\doi{10.1007/JHEP07(2013)116.}}}

\begin{document}\cmsNoteHeader{FSQ-12-010}

\hyphenation{had-ron-i-za-tion}
\hyphenation{cal-or-i-me-ter}
\hyphenation{de-vices}

\RCS$Revision: 191938 $
\RCS$HeadURL: svn+ssh://alverson@svn.cern.ch/reps/tdr2/papers/FSQ-12-010/trunk/FSQ-12-010.tex $
\RCS$Id: FSQ-12-010.tex 191938 2013-06-23 12:19:49Z alverson $
\newlength\cmsFigWidth
\ifthenelse{\boolean{cms@external}}{\setlength\cmsFigWidth{0.85\columnwidth}}{\setlength\cmsFigWidth{0.4\textwidth}}
\ifthenelse{\boolean{cms@external}}{\providecommand{\cmsLeft}{top}}{\providecommand{\cmsLeft}{left}}
\ifthenelse{\boolean{cms@external}}{\providecommand{\cmsRight}{bottom}}{\providecommand{\cmsRight}{right}}
\cmsNoteHeader{FSQ-12-010} % This is over-written in the CMS environment: useful as preprint no. for export versions
\title{Study of exclusive two-photon production of \texorpdfstring{$\PWp\PWm$ in $\Pp\Pp$}{W(+)W(-) in pp} collisions at $\sqrt{s}=7\TeV$ and constraints on anomalous quartic gauge couplings}

\date{\today}

\abstract{
A search for exclusive or quasi-exclusive $\PWp\PWm$ production by photon-photon interactions, $\Pp\Pp \rightarrow\Pp^{(*)}\PWp\PWm\Pp^{(*)}$, at
$\sqrt{s}=7\TeV$ is reported using data collected by the CMS detector with an integrated luminosity of 5.05\fbinv. Events
are selected by requiring a $\mu^{\pm}\Pe^{\mp}$ vertex with no additional associated charged tracks and dilepton transverse momentum $\pt(\mu^{\pm}\Pe^{\mp})>30\GeV$.
Two events passing all selection requirements are observed in the data, compared to a standard model expectation of $2.2\pm0.4$ signal events
with $0.84\pm0.15$ background. The tail of the dilepton $\pt$ distribution is studied for deviations from the standard
model. No events are observed with $\pt > 100$\GeV. Model-independent upper limits are computed and compared to predictions involving anomalous quartic
gauge couplings. The limits on the parameters $a^{\PW}_{0,C}/\Lambda^{2}$ with a dipole form factor and an energy cutoff $\Lambda_{\text{cutoff}}=500\GeV$
are of the order of $10^{-4}$.}

\hypersetup{%
pdfauthor={CMS Collaboration},%
pdftitle={Study of exclusive two-photon production of W(+)W(-) in pp collisions at sqrt(s)=7 TeV and constraints on anomalous quartic gauge couplings},%
pdfsubject={CMS},%
pdfkeywords={CMS, physics, gauge couplings}}

\maketitle %maketitle comes after all the front information has been supplied

\section{Introduction}

The detection of high-energy photon interactions at the Large Hadron Collider (LHC) opens up the possibility of interesting and novel
research \cite{Enterria:2008zz,deFavereaudeJeneret:2009db}. In particular, measurements of the two-photon production of a pair of $\PW$-bosons provide sensitivity to anomalous quartic gauge couplings of the gauge bosons. Exploratory
studies \cite{Pierzchala:2008xc,Chapon:2009hh} showed potential for extending the experimental reach by several orders of magnitude with respect to the best
limits so far obtained at the Tevatron~\cite{Abazov:2013opa} and
LEP~\cite{Belanger:1999aw,Heister:2004yd,Abbiendi:2004bf,Abbiendi:2003jh,Abbiendi:1999aa,Achard:2002iz,Achard:2001eg,Abdallah:2003xn}.
First measurements of the exclusive two-photon production of muon and electron pairs at $\sqrt{s}=7\TeV$, $\Pp\Pp \to \Pp \ell^{+}\ell^{-} \Pp$, were made using
$\sim$40\pbinv of data collected with the Compact Muon Solenoid (CMS) at the LHC in 2010~\cite{Chatrchyan:2011ci,Chatrchyan:2012tv}. The present
analysis is based on the experimental technique developed in Ref.~\cite{Chatrchyan:2011ci} and
uses the full data sample collected by the CMS experiment in 2011.

In this analysis the $\mu^{\pm}\Pe^{\mp}$ final state is used to search for fully exclusive (``elastic'') $\Pp\Pp \rightarrow \Pp\PWp\PWm\Pp$ production.
Since both very forward-scattered protons escape detection, such a production process is characterized by a primary vertex formed from a
$\mu^{\pm}\Pe^{\mp}$ pair with no other tracks, with large transverse momentum, $\pt(\mu^{\pm}\Pe^{\mp})$, and large invariant mass, $m(\mu^{\pm}\Pe^{\mp})$. This
signature is also accessible via quasi-exclusive (``inelastic'' or ``proton dissociative'') production, in which one or both of the incident
protons dissociate into a low-mass system that escapes detection, denoted as $\Pp^{*}$. The two-photon signal $\gamma\gamma\rightarrow \PWp\PWm$ is
therefore comprised of both the elastic and inelastic contributions.

In the case of decays of the $\PWp\PWm$ pair to same-flavor $\Pgmp\Pgmm$ or $\Pep\Pem$ final states, the backgrounds are more
than an order of magnitude larger than in the $\mu^{\pm}\Pe^{\mp}$ final state. Therefore in the present analysis, only the
$\mu^{\pm}\Pe^{\mp}$ channel is used to search for a $\Pp\Pp \rightarrow\Pp^{(*)}\PWp\PWm\Pp^{(*)}$ signal. We use the
$\Pgmp\Pgmm$ channel to select a control sample of high-mass $\Pp\Pp \rightarrow\Pp^{(*)}\Pgmp\Pgmm\Pp^{(*)}$ events originating
mainly from direct $\gamma\gamma\rightarrow\Pgmp\Pgmm$ production. Final states containing a $\mu^{\pm}\Pe^{\mp}$ pair may arise from direct decays
of $\PW^\pm$ bosons to electrons and muons or from $\PW\rightarrow\tau\nu$ decays, with the $\tau$ subsequently decaying to an electron or a muon. For
brevity, we will refer to the full reaction as $\Pp\Pp \rightarrow\Pp^{(*)}\PWp\PWm\Pp^{(*)} \rightarrow\Pp^{(*)}\mu^{\pm}\Pe^{\mp}\Pp^{(*)}$,
where the final state is understood to contain between two and four undetected neutrinos, in addition to the charged $\mu^{\pm}\Pe^{\mp}$
pair.

We first use the $\Pp\Pp \rightarrow\Pp^{(*)}\Pgmp\Pgmm\Pp^{(*)}$ control sample to validate the selection by comparing
the expected and observed numbers of events and to estimate from the data the proton dissociative contribution. The
dominant backgrounds in the $\mu^{\pm}\Pe^{\mp}$ channel, due to the inclusive production of $\PWp\PWm$ and $\tau^{+}\tau^{-}$ pairs, are then
constrained using control regions with low $\pt(\mu^{\pm}\Pe^{\mp})$ or a low-multiplicity requirement for extra tracks originating from the $\mu^{\pm}\Pe^{\mp}$ vertex.

The data for the signal region are then compared to the standard model (SM) expectation for the backgrounds and the
$\gamma\gamma\rightarrow \PWp\PWm$ signal. Finally, tails of the $\pt(\mu^{\pm}\Pe^{\mp})$ distribution, where the
SM $\gamma\gamma\rightarrow \PWp\PWm$ contribution is expected to be small, are investigated to look for
anomalous quartic gauge couplings~\cite{Belanger:1992qh}.

The paper is organized as follows. Section~\ref{cms-detector} describes the CMS detector. Section~\ref{theory} presents the theory related to the
$\gamma\gamma\rightarrow \PWp\PWm$ process and the approach to deal with the anomalous couplings, together with the description of the simulated
data. In Section~\ref{event-selection} the trigger, lepton identification, and preselection criteria employed in this analysis are presented in detail.
The study of the $\Pp\Pp \rightarrow\Pp^{(*)}\Pgmp\Pgmm\Pp^{(*)}$ control sample is discussed in Section~\ref{mumu}, while the investigation of the $\mu^\pm \Pe^\mp$
signal for the elastic and proton dissociative $\PWp\PWm$ production is presented in Section~\ref{mue}. Next, Section~\ref{systematics} describes the impact
of the systematic uncertainties encountered in this analysis along with the additional checks for the background modelling. Finally, the results of this
analysis are presented in Section~\ref{results}, followed by the summary in Section~\ref{summary}.

\section{The CMS detector}
\label{cms-detector}

A detailed description of the CMS experiment can be found elsewhere~\cite{JINST}. The central feature of the CMS apparatus is a
superconducting solenoid, of 6\unit{m} internal diameter.  Within the field volume are the silicon pixel and strip tracker, the crystal electromagnetic
calorimeter (ECAL), and the brass-scintillator hadron calorimeter (HCAL).  Muons are measured in gas-ionization detectors embedded in the steel flux-return
yoke of the magnet. Besides the barrel and endcap detectors, CMS has extensive forward calorimetry.

The CMS experiment uses a right-handed coordinate system, with the origin at the nominal collision point, the $x$ axis pointing towards the center of the
LHC ring, the $y$ axis pointing up (perpendicular to the plane of LHC ring), and the $z$ axis along the anticlockwise-beam direction. The polar angle
$\theta$ is measured from the positive $z$ axis and the azimuthal angle $\phi$ is measured, in radians, in the $(x,y)$ plane relative to the $x$ axis. The
silicon tracker covers a range of $\abs{\eta}<2.4$, where $\eta = -\ln[\tan(\theta/2)]$, and consists of three layers made of 66~million $100\times150\mum^{2}$ pixels followed by ten microstrip
layers, with strips of pitch between 80 and 180\mum. Muons are measured in the $\abs{\eta}<2.4$ range, with detection planes made of three technologies: drift
tubes, cathode strip chambers, and resistive-plate chambers. The 3.8\unit{T} magnetic field, and the high granularity of the silicon
tracker, allow the transverse momentum of the muons matched to tracks in the silicon detector to be measured with a resolution better than $\sim$1.5\% for \pt smaller than 100\GeV.
The ECAL provides coverage in a range of $\abs{\eta}<1.479$ in the barrel region and $1.479<\abs{\eta}<3.0$ in the two endcap regions.
The first level of the CMS trigger system, composed of custom hardware processors, uses information from the calorimeters and muon detectors to
select (in less than 3\mus) the most interesting events. The high-level trigger (HLT) processor farm further decreases the event
rate from 100\unit{kHz} to a few hundred Hz before data storage.

\section{Theory and simulation}
\label{theory}

\begin{figure}[h!t]
\centering

\begin{minipage}[h]{0.25\textwidth}
\center{\includegraphics[width=.9\textwidth]{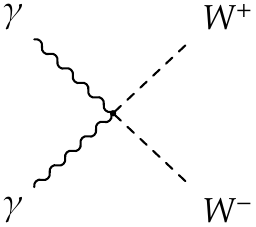}} \\ (a)
\end{minipage}
\hspace{.2em}
\begin{minipage}[h]{0.25\textwidth}
\center{\includegraphics[width=.9\textwidth]{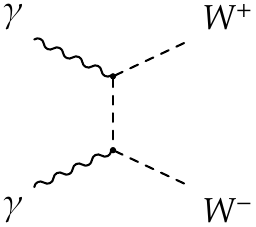}} \\ (b)
\end{minipage}
\hspace{.2em}
\begin{minipage}[h]{0.25\textwidth}
\center{\includegraphics[width=.9\textwidth]{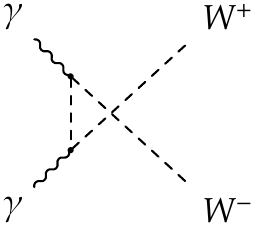}} \\ (c)
\end{minipage}

\caption{\small{Quartic gauge coupling (a) and $t$- (b) and $u$-channel (c) $\PW$-boson exchange diagrams contributing to the $\gamma\gamma\to\PWp\PWm$ process at leading
order in the SM.}}
\label{fig:FeynmanDiagrams}
\end{figure}

The electroweak sector of the SM~\cite{Glashow:1961tr,Weinberg:1967tq,sm_salam} predicts 3- and 4-point vertices with the gauge bosons, which are
represented in the SM Lagrangian by the following terms for the quartic $\PW\PW\gamma\gamma$ and triple $\PW\PW\gamma$ couplings:
\begin{equation}
\begin{aligned}
L^{\PW\PW\gamma}       &= -ie    \left( \partial_{\mu}A_{\nu} - \partial_{\nu}A_{\mu} \right) W^{+\mu}W^{-\nu} \\
L^{\PW\PW\gamma\gamma} &= -e^{2} \left( W^{+}_{\mu}W^{-\mu}A_{\nu}A^{\nu} - W^{+}_{\mu}W^{-}_{\nu}A^{\mu}A^{\nu} \right)
\end{aligned}
\end{equation}

where $A^{\mu}$ is the photon field and $W^{\mu}$ is the $\PW$-boson field. As a result, the diagrams that represent the $\PW\PW\gamma\gamma$
interaction at lowest order in the perturbation series consist of both quartic gauge coupling (Fig.~\ref{fig:FeynmanDiagrams}(a)) and $t$- and
$u$-channel $\PW$-boson exchange diagrams (Fig.~\ref{fig:FeynmanDiagrams}(b,c)).

Measurements of the quartic $\PW\PW\gamma\gamma$ coupling can be used to look for any deviation from the SM predictions, which would reveal a sign of
new physics~\cite{Belanger:1999aw}. One has to take into account more generic couplings in order to study the possibility of such deviations
in high-energy collisions. Considering models with the anomalous triple gauge couplings, the quartic $\PW\PW\gamma\gamma$ and triple $\PW\PW\gamma$
couplings can be associated with a single anomalous dimension-six operator~\cite{Belanger:1992qh}. The genuine anomalous quartic gauge couplings
considered here are instead
introduced via an effective Lagrangian containing new terms respecting local $U(1)_\mathrm{EM}$ and global custodial $SU(2)_{C}$ symmetry. Further
imposing charge-conjugation and parity symmetries, $C$- and $P$, results in a minimum of two additional dimension-six terms, containing the parameters $a^{\PW}_{0}$ and
$a^{\PW}_{C}$~\cite{Belanger:1992qh}:
\begin{equation}
\begin{aligned}
L^{0}_{6} &= \frac{e^{2}}{8}\frac{a^{\PW}_{0}}{\Lambda^{2}}F_{\mu\nu}F^{\mu\nu}W^{+\alpha}W^{-}_{\alpha}-\frac{e^{2}}{16\cos^{2}\Theta_{\PW}}\frac{a^{Z}_{0}}{\Lambda^{2}}F_{\mu\nu}F^{\mu\nu}Z^{\alpha}Z_{\alpha} \\
L^{C}_{6} &= \frac{-e^{2}}{16}\frac{a^{\PW}_{C}}{\Lambda^{2}}F_{\mu\alpha}F^{\mu\beta}(W^{+\alpha}W^{-}_{\beta}+W^{-\alpha}W^{+}_{\beta})-\frac{e^{2}}{16\cos^{2}\Theta_{\PW}}\frac{a^{Z}_{C}}{\Lambda^{2}}F_{\mu\alpha}F^{\mu\beta}Z^{\alpha}Z_{\beta}
\end{aligned}
\end{equation}
where $\Lambda_{\text{cutoff}}$ is the energy cutoff scale for the form factor and the second terms in the expressions are those corresponding to
the $\cPZ$-boson couplings. These genuine anomalous quartic couplings are therefore completely independent of the SM triple and quartic gauge couplings.
While the $\gamma\gamma \rightarrow \PWp\PWm$ process contains two triple gauge coupling vertices
involving t-channel $\PW$-boson exchange (Fig.~\ref{fig:FeynmanDiagrams} left), the sensitivity to anomalous triple gauge
couplings is not expected to significantly surpass the existing experimental limits on $\PW\PW\gamma$ couplings from single triple gauge coupling
processes~\cite{Chapon:2009hh}. Hence, only anomalous quartic gauge couplings are considered in the analysis, assuming no anomalous triple gauge
couplings are present.

The existing constraints on anomalous quartic gauge couplings from $\Pep\Pem$ collisions at LEP are derived
from $\Pep\Pem\rightarrow \PWp\PWm \gamma$ and $\PWp\PWm \rightarrow \gamma\gamma$ interactions in
which the effective center-of-mass energy is limited to values well below the $\Pep\Pem$ center-of-mass
energy of $\sqrt{s}=209\GeV$. In contrast, the spectrum of $\gamma\gamma$ interactions at
the LHC and the Tevatron extends to much higher values, resulting in increased sensitivity to anomalous couplings.

The $\gamma\gamma~\to~\PWp\PWm$ cross section increases quadratically with anomalous coupling strength, and consequently unitarity
is violated for high-energy $\gamma\gamma$ interactions. For anomalous couplings $a^{\PW}_{0}/\Lambda^{2}$, $a^{\PW}_{C}/\Lambda^{2}$ of order $10^{-5}$, the
unitarity bound is already reached for collisions with a $\gamma\gamma$ center-of-mass energy $\PW_{\gamma\gamma} \sim 1\TeV$~\cite{Pierzchala:2008xc, Chapon:2009hh}. In order to tame this
rising of the cross section, both $a^{\PW}_{0}/\Lambda^{2}$ and $a^{\PW}_{C}/\Lambda^{2}$ parameters are multiplied by a form factor :
\begin{equation*}
a^{\PW}_{0,C}(W^{2}_{\gamma\gamma}) = \frac{a^{\PW}_{0,C}}{\left( 1+\frac{W^{2}_{\gamma\gamma}}{\Lambda_{\text{cutoff}}^{2}} \right)^{p}}
\end{equation*}
where $p$ is a free parameter,
which is conventionally set to 2 (dipole form factor), following previous studies of anomalous quartic gauge couplings~\cite{Abbott:1999ec,Abazov:2006xq}. Because
the new physics that enters to regulate the cross section has an energy scale $\Lambda$ and a form that are a priori unknown,
we consider both a scenario with a dipole form factor with energy cutoff scale $\Lambda_{\text{cutoff}}=500\GeV$, and a scenario with no form factor (i.e.,
$\Lambda_{\text{cutoff}}\to\infty$).

The $\gamma\gamma\rightarrow \PWp\PWm$ signal is generated using {\CALCHEP{2.5.4}}~\cite{Pukhov:2004ca},
with {\PYTHIA{6.422}}~\cite{Sjostrand:2006za} used to simulate the decay of the $\PWp\PWm$ pair. The simulated inclusive background samples
used in this analysis are produced with {\MADGRAPH{5}}~\cite{Alwall:2011uj} for $\PWp\PWm + \text{jets}$,
$\PW$ + jets, and $t\bar{t}$ processes, and with {\textsc{powheg 1.0}}~\cite{Nason:2004rx,Frixione:2007vw,Alioli:2010xd,Sjostrand:2006za}
for $\tau^{+}\tau^{-}$ pairs produced via the Drell--Yan process. In the simulated $\PWp\PWm + \text{jets}$, $\PW + \text{jets}$, $\ttbar$, and Drell--Yan background
samples, $\tau$-decays are handled by the \TAUOLA~\cite{Jadach:1993hs} package. Inclusive $\PWp\PWm$ production at the LHC is expected to be dominated by
$s$-channel and $t$-channel $q\overline{q} \rightarrow \PWp\PWm$ interactions, followed by a small ($\sim3\%$) contribution from gluon-gluon
interactions~\cite{Chatrchyan:2013yaa}. This inclusive $\PWp\PWm$ background sample is scaled to the next-to-leading-order (NLO)
prediction obtained from \MCFM~\cite{Campbell:2010ff}, which describes the experimentally measured cross section~\cite{Chatrchyan:2011tz,ATLASWW} within
the uncertainties. The underlying event in all background processes is simulated using the Z2 tune~\cite{Field:2011iq} of {\PYTHIA}.

In addition to the inclusive $\PWp\PWm$ backgrounds, we consider $\PWp\PWm$ production from single diffractive interactions, and from
$\PW\PW \rightarrow \PW\PW$ scattering (vector boson fusion). The diffractive $\PWp\PWm$ background is generated using \textsc{pompyt}~\cite{Bruni:1993is}.
Single diffractive $\PWp\PWm$ production will result in events with a multiplicity of extra tracks lower than that of non-diffractive production, and with
large theoretical uncertainties related to survival probabilities that will suppress the visible cross section. We conservatively consider the
diffractive $\PWp\PWm$ background with no survival probability correction in the default background estimate. The two-photon processes,
$\gamma\gamma\rightarrow\Pgmp\Pgmm$ and $\gamma\gamma\rightarrow\tau^{+}\tau^{-}$, are produced using {\textsc{lpair}}~\cite{Vermaseren:1982cz,Baranov:1991yq},
which describes well the exclusive and quasi-exclusive dilepton measurements of CMS~\cite{Chatrchyan:2011ci,Chatrchyan:2012tv}. The contribution
from (gluon-mediated) central exclusive $\PWp\PWm$ production is estimated to be ${\lesssim}1\%$ of the $\gamma\gamma \rightarrow \PWp\PWm$ cross
section~\cite{Lebiedowicz:2012gg} and is neglected in the current analysis. The {\textsc{vbfnlo}} generator~\cite{Arnold:2011wj} is used to study
backgrounds from $\PW\PW \rightarrow \PW\PW$ scattering, with {\PYTHIA} used for hadronization and the decay of the $\PWp\PWm$ pair. All signal and
background samples are produced with a detailed {\GEANTfour~\cite{Agostinelli:2002hh} simulation of the CMS detector.

\section{Event selection}
\label{event-selection}

The data used in this analysis correspond to the full sample collected in 2011 at $\sqrt{s}=7\TeV$ with the CMS detector.
In the $\mu^{\pm}\Pe^{\mp}$ channel all detector subsystems are required to pass the standard data quality requirements, resulting in a sample with an integrated luminosity of 5.05\fbinv. In the $\Pgmp\Pgmm$ channel, which is used only as a control sample, a less restrictive selection is used, requiring that
only the muon or tracking systems pass the data quality requirements. This results in a sample with a slightly higher integrated luminosity of
5.24\fbinv.

In the $\mu^{\pm}\Pe^{\mp}$ channel events are selected online by two electron-muon HLT algorithms with asymmetric thresholds. The first
algorithm requires a muon of 17\GeV and an electron of 8\GeV, while the second requires a muon of 8\GeV and an electron of 17\GeV.
In the $\mu^{\pm}\mu^{\mp}$ channel, dimuon triggers with asymmetric 17\GeV and 8\GeV thresholds on the two muons are used for
consistency with the $\mu^{\pm}\Pe^{\mp}$ channel.

Muon candidates are required to pass a tight muon selection similar to that described in detail in Ref.~\cite{Chatrchyan:2011ci}. Electrons are required to pass
a medium identification selection with criteria chosen to ensure the offline selection is tighter than the trigger requirement. The electron selection is similar to
that of Ref.~\cite{CMS:2010bta} and includes requirements on the shower shape measured in the ECAL, the transverse and longitudinal impact parameters with
respect to the primary vertex, an isolation criterion based on combined information from the silicon tracker and calorimeters, the number of missing
hits on the electron track, and the incompatibility of the electron and nearby tracks with those originating from a photon conversion. A particle
flow (PF) algorithm~\cite{CMS:2009nxa} is used to reconstruct particles in the event by combining information from all detector systems. The missing
transverse energy $\ETslash$ is then computed from the negative vector sum of all particles.

After trigger selection and lepton identification, a first preselection criterion is applied offline on the
data by requiring a reconstructed muon and electron of opposite charge, each having $\pt >20\GeV$ and $\abs{\eta}< 2.4$,
matched to a common primary vertex with fewer than 15 additional tracks. After the trigger selection, the leptons are required to have an invariant mass
$m(\ell^{+}\ell'^{-})>20\GeV$ in both the $\mu^{\pm}\Pe^{\mp}$ and $\Pgmp\Pgmm$ channels. In the remainder of this paper we will use the notation
$\pt(\ell^{+},\ell'^{-})$ to indicate a $\pt$ selection applied to each lepton of the pair, and $\pt(\ell^{+}\ell'^{-})$ to indicate the $\pt$ of
the pair.

At high luminosities, almost all signal events will have additional interactions within the same bunch crossing (``pileup''), that
produce extra charged tracks and extra activity in the calorimeters. During the 2011 LHC run the average number of interactions per
crossing was approximately 9. In order to retain efficiency in high pileup conditions, a selection based only on the number of charged tracks
originating from the same primary vertex as the $\ell^{+}\ell'^{-}$ pair is used, similar to the method in Ref.~\cite{Chatrchyan:2011ci}.

In the $\mu^{\pm}\Pe^{\mp}$ channel the SM signal region is defined to have zero extra tracks associated with
the $\mu^{\pm}\Pe^{\mp}$ vertex, and transverse momentum of the pair $\pt(\mu^{\pm}\Pe^{\mp})>30\GeV$. The first requirement rejects backgrounds
from inclusive production, while the second is chosen based on the simulated $\pt(\mu^{\pm}\Pe^{\mp})$ distribution of the signal and
$\tau^{+}\tau^{-}$ background events. The efficiency for reconstruction of primary vertices with two or more tracks has been measured to
be $\geq$98\% in simulation, and $\geq$99\% in data. In addition, events are only accepted as $\mu^{\pm}\Pe^{\mp}$ events if they have failed to
satisfy the $\mu^{\pm}\mu^{\mp}$ selection, in order to reject $\gamma\gamma\rightarrow\Pgmp\Pgmm$ events with the muon misidentified as an
electron due to a bremsstrahlung photon overlapping with the muon track.

For the anomalous quartic gauge couplings search, a restricted region of $\pt(\mu^{\pm}\Pe^{\mp})>100\GeV$ is used. This is chosen to reduce the expected
SM $\gamma\gamma\rightarrow \PWp\PWm$ contribution to  approximately 0.1~events after all selection requirements, while retaining
sensitivity to anomalous couplings of order $10^{-4}$ for $\Lambda_{\text{cutoff}}=500\GeV$ or larger (Fig.~\ref{fig:aqgc_no-cuts}). This corresponds to values
of the anomalous quartic gauge coupling parameters approximately two orders of magnitude smaller than the limits obtained at
LEP~\cite{Abbiendi:2004bf,Achard:2002iz,Abdallah:2003xn} and approximately one order of magnitude smaller than Tevatron limits~\cite{Abazov:2013opa}.

\begin{figure}[h!t]
\centering
\includegraphics[width=.75\textwidth]{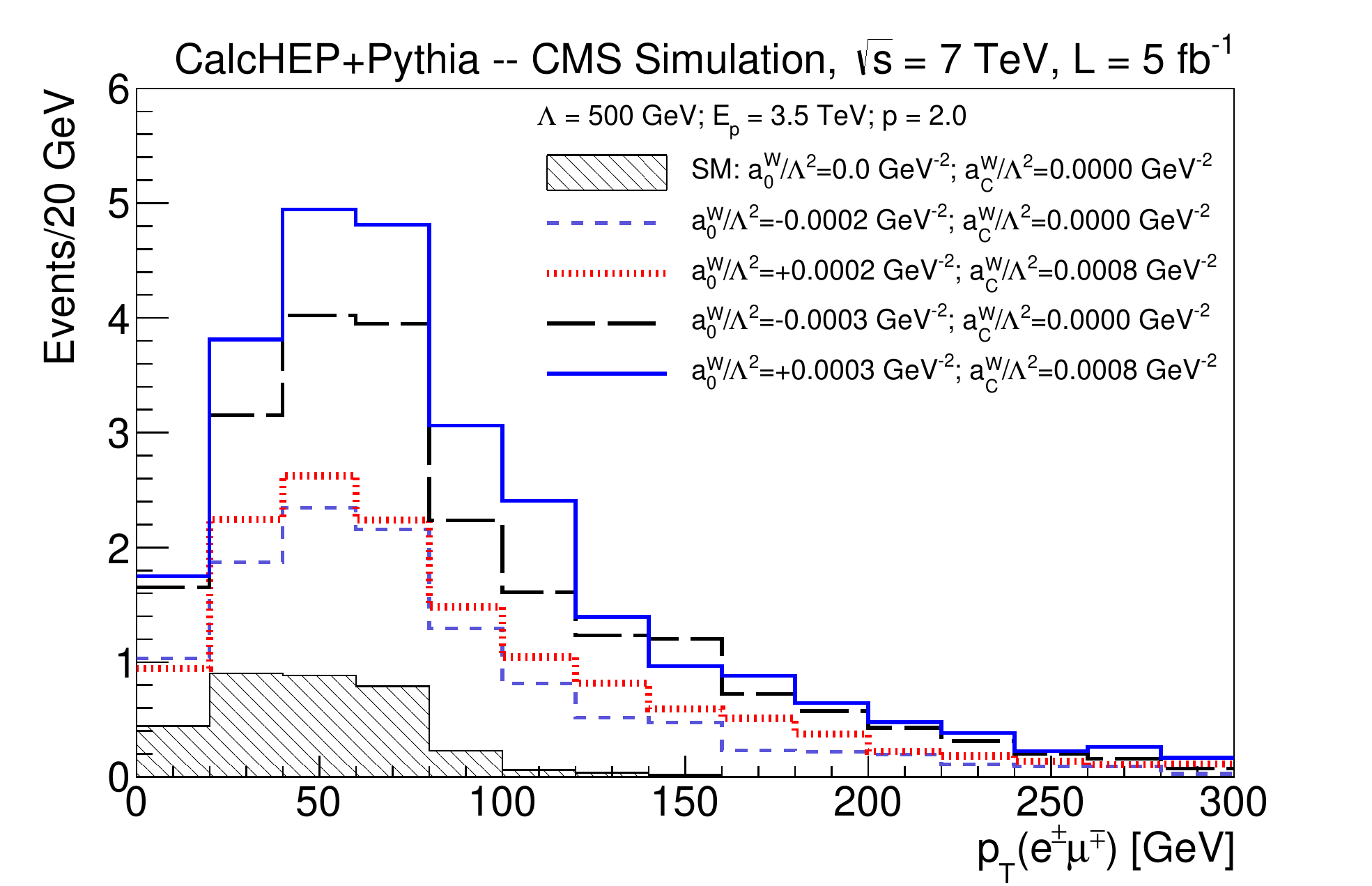}
\caption{\small{Transverse momentum distribution of lepton pairs for the $\gamma\gamma\to \PWp\PWm$ process at generator level in the SM (shaded histogram),
and for several values of anomalous quartic gauge coupling parameters $a^{\PW}_{0}$ and $a^{\PW}_{C}$ (open histograms). In this plot
$\Lambda_{\text{cutoff}}=500\GeV$ is the energy cutoff scale in the dipole form factor.
}}
\label{fig:aqgc_no-cuts}
\end{figure}

\section{Cross checks with \texorpdfstring{$\Pgmp\Pgmm$}{mu(+)mu(-)} events}\label{sect:mumu-splitted-runepochs}
\label{mumu}

In the case of same-flavor dilepton final states, the background due to Drell--Yan ($\Pp\Pp \rightarrow \ell^{\pm}\ell^{\mp}X$) or
$\Pp\Pp \rightarrow\Pp^{(*)}\ell^{\pm}\ell^{\mp}\Pp^{(*)}$ production is more than an order of magnitude larger than the unlike-flavor
$\mu^{\pm}\Pe^{\mp}$ channel of the $\gamma\gamma\rightarrow \PWp\PWm$ signal for the same selection criteria. The semileptonic
and fully hadronic channels, in which one or both $\PW$s decay into a $\cPq\cPaq$ pair, are similarly dominated by background
and complicated by the requirement of matching the constituents of the resulting jets to the primary vertex. As a test benchmark
for high-mass lepton pair detection, elastic $\gamma\gamma \rightarrow \Pgmp\Pgmm$ production is used because of the
small theoretical uncertainties on the cross section~\cite{Vermaseren:1982cz}.

The dimuon sample with zero extra tracks is divided into two kinematic regions based
on the \pt balance ($\abs{\Delta\pt(\Pgmp\Pgmm)}$) and acoplanarity ($1-\abs{\Delta\phi(\Pgmp\Pgmm)/\pi}$) of the pair.
The first region with $1-\abs{\Delta\phi(\Pgmp\Pgmm)/\pi}<0.1$ and $\abs{\Delta\pt(\Pgmp\Pgmm)}<1\GeV$ is defined as the ``elastic" region, where the
dimuon kinematic requirements are consistent with elastic $\Pp\Pp \rightarrow \Pp \Pgmp\Pgmm \Pp$ events where both protons remain intact~\cite{Chatrchyan:2011ci}.
The second region with $1-\abs{\Delta\phi(\Pgmp\Pgmm)/\pi}>0.1$ or $\abs{\Delta\pt(\Pgmp\Pgmm)}>1\GeV$ (``dissociation'' selection) is dominated by $\gamma\gamma\rightarrow\Pgmp\Pgmm$
interactions in which one or both protons dissociate. The latter process is less well-known theoretically, and subject to corrections from
rescattering, in which strong interactions between the protons produce additional hadronic activity. As this effect is not included in the simulation,
it may lead to a significant over-estimate of the proton dissociation contribution in two-photon interactions~\cite{HarlandLang:2012qz}. We
therefore use this second control region to estimate the proton dissociation yield directly from data.

The contributions from exclusive $\cPZ$ production are expected to be negligible compared to the cross section of approximately 0.5\unit{pb} from
$\gamma\gamma\rightarrow\Pgmp\Pgmm$, with $\pt(\Pgmp,\Pgmm)>20$\GeV. Exclusive $\cPZ$ photoproduction, $\gamma \Pp \rightarrow \cPZ \Pp$, is expected to have
a cross section smaller than 1\unit{fb} after taking into account the branching fraction to $\Pgmp\Pgmm$~\cite{Goncalves:2007vi,Motyka:2008ac,Cisek:2009hp},
while the $\gamma\gamma \rightarrow \cPZ$ process is forbidden at tree level. The $\cPZ$-boson peak therefore provides another cross-check
of the residual inclusive Drell--Yan contamination in both regions. In Fig.~\ref{fig:mumu-lumi-runab-invm} the invariant mass distribution in the elastic-enhanced
region is shown, with the marked $\cPZ$-peak region defined
as $70\GeV<m(\Pgmp\Pgmm)<106\GeV$. In Fig.~\ref{fig:mumu-lumi-runab} the dimuon kinematic distributions for events having zero
extra tracks are shown. The distributions are plotted separately for the $\cPZ$-peak region, which is expected to include a large inclusive
Drell--Yan component, and for the region outside the $\cPZ$ peak, which is expected to be dominated by two-photon interactions. For the
kinematic distributions with zero extra tracks originating from the $\Pgmp\Pgmm$ production vertex, good agreement is observed between data and simulation. This
confirms that pileup effects and low-multiplicity fluctuations of the inclusive Drell--Yan processes are well modeled. The hatched bands indicate the
statistical uncertainty. In Fig.~\ref{fig:mumu-alumi-invmass} the dimuon pair invariant mass is plotted for the dissociation selection with zero extra tracks.

\begin{figure}[h!t]
\centering
\includegraphics[width=.8\textwidth]{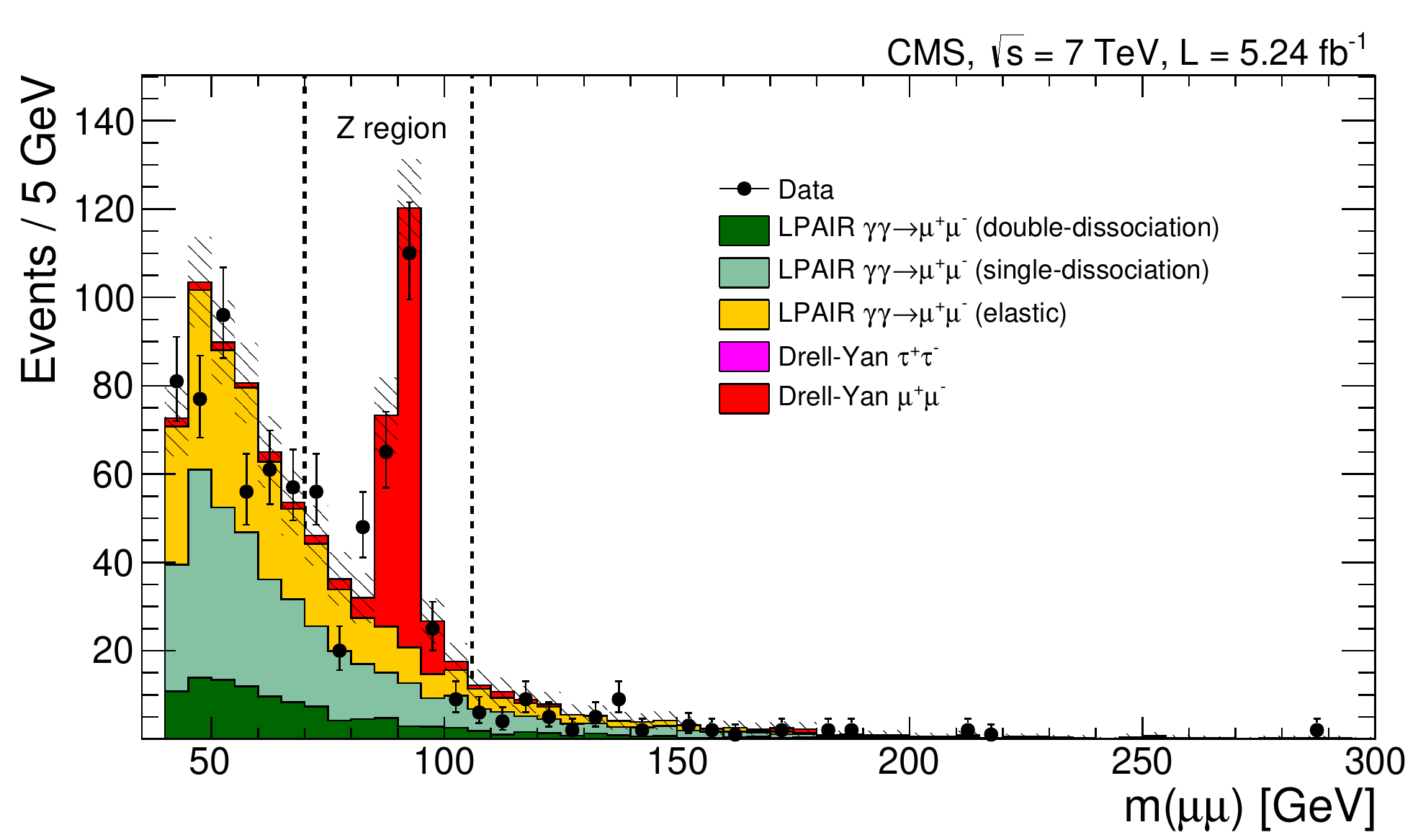}
\caption{\small{Invariant mass distribution of the muon pairs for the elastic selection with no additional track on the dimuon vertex. The dashed lines indicate the $\cPZ$-peak region. The hatched bands indicate the statistical uncertainty in the simulation.}}
\label{fig:mumu-lumi-runab-invm}
\end{figure}

\begin{figure}[h!t]
\centering
\includegraphics[width=6.5cm]{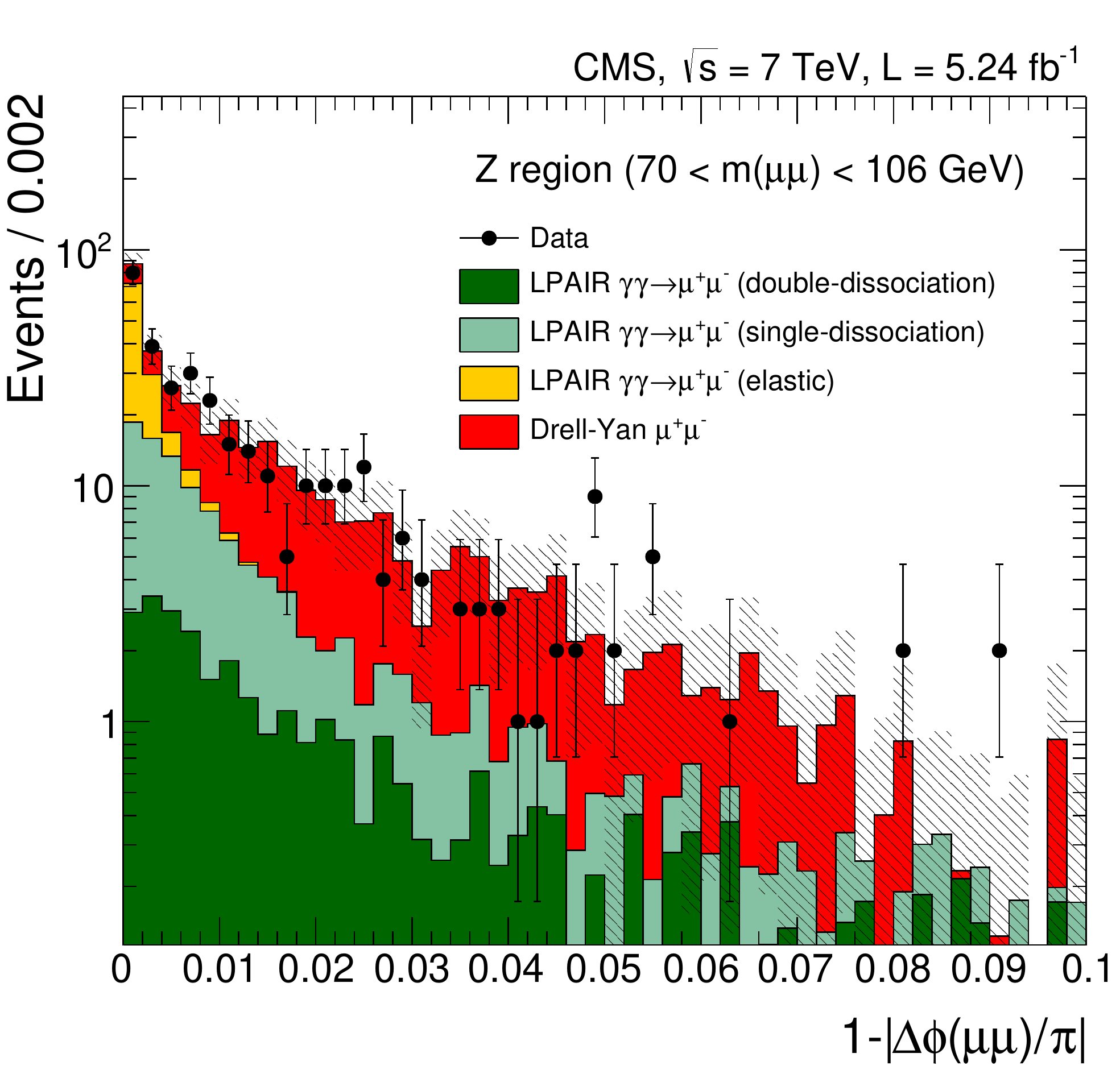}
\includegraphics[width=6.5cm]{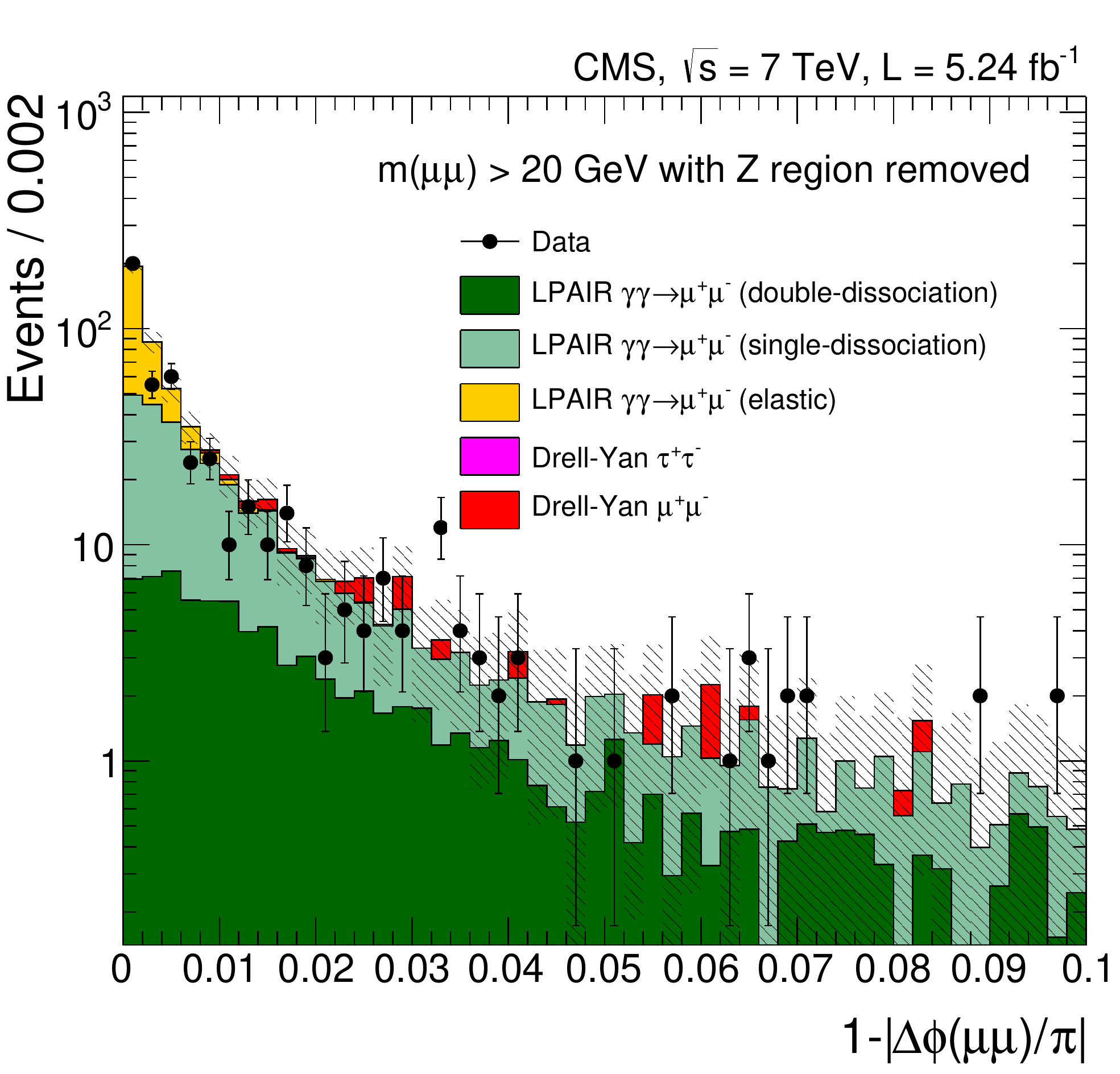}
\includegraphics[width=6.5cm]{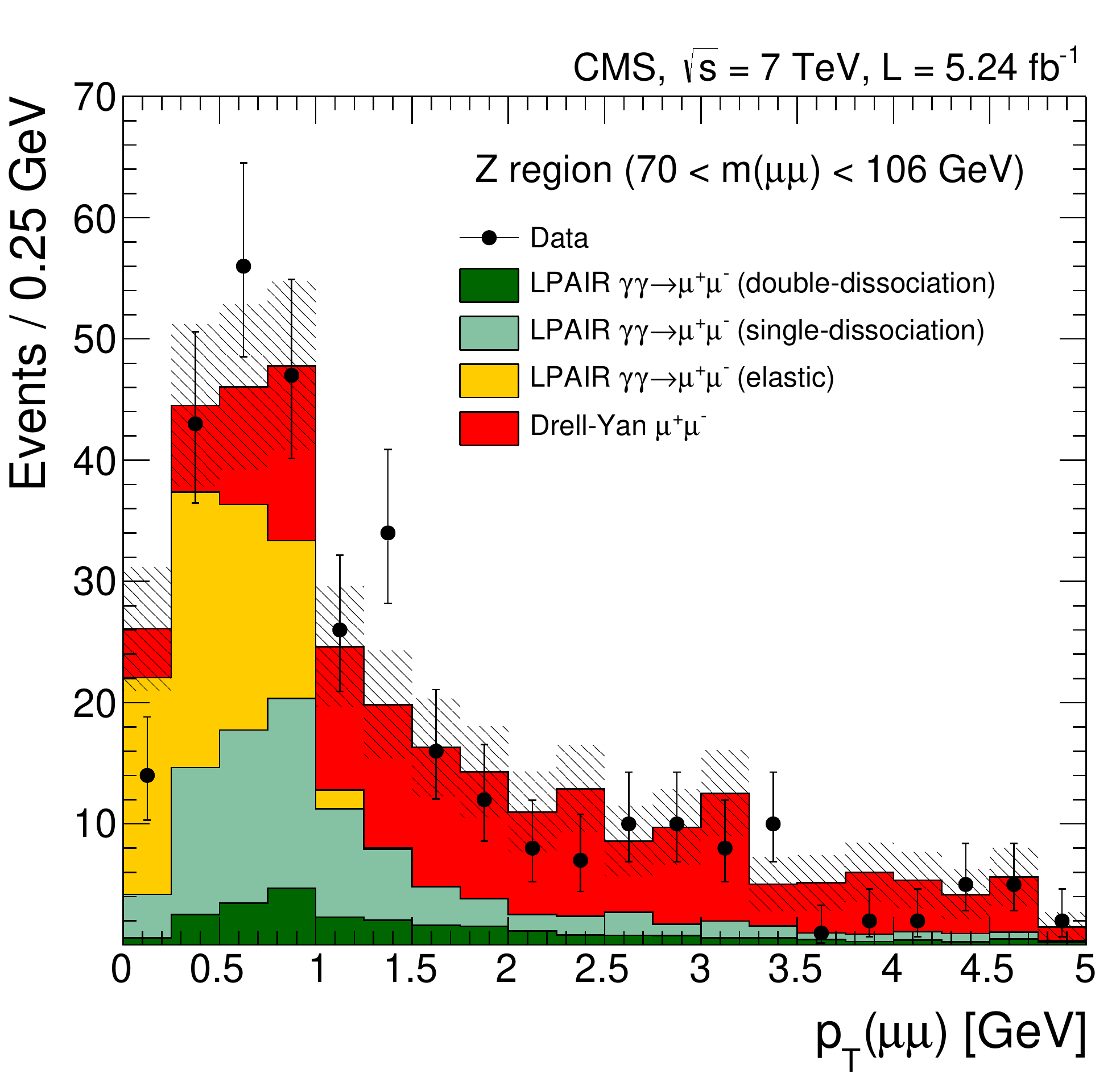}
\includegraphics[width=6.5cm]{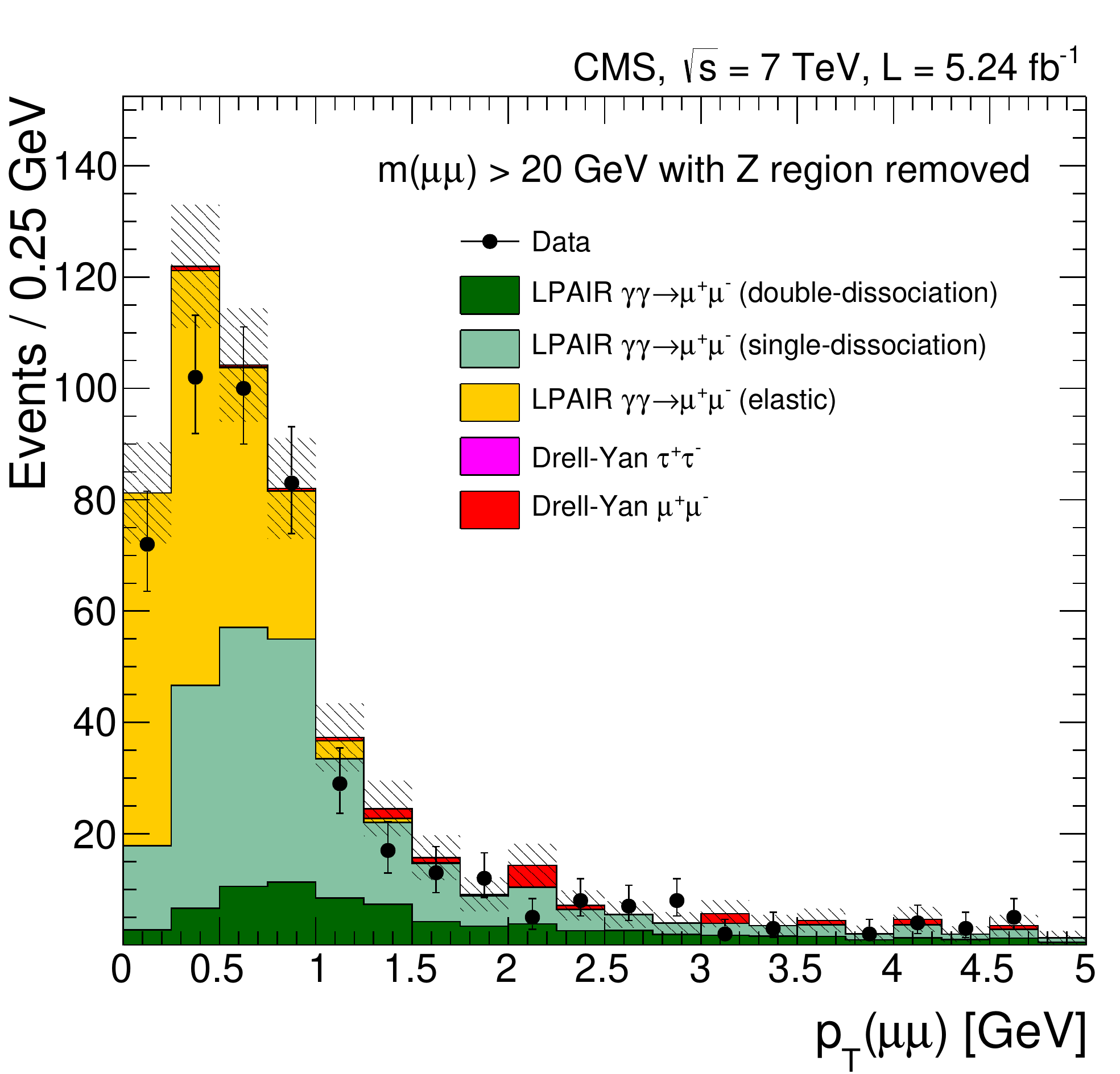}
\caption{\small{Kinematic distributions for the elastic selection, for the $\cPZ$ region only ($70\GeV{<m(\Pgmp\Pgmm)<}106\GeV$, left panel) and with the $\cPZ$
region removed (right panel). The acoplanarity (above) and $\pt$ of $\Pgmp\Pgmm$ pairs with zero extra tracks (below) are shown. The hatched bands
indicate the statistical uncertainty in the simulation.}}
\label{fig:mumu-lumi-runab}
\end{figure}

\begin{figure}[h!t]
\centering
\includegraphics[width=.8\textwidth]{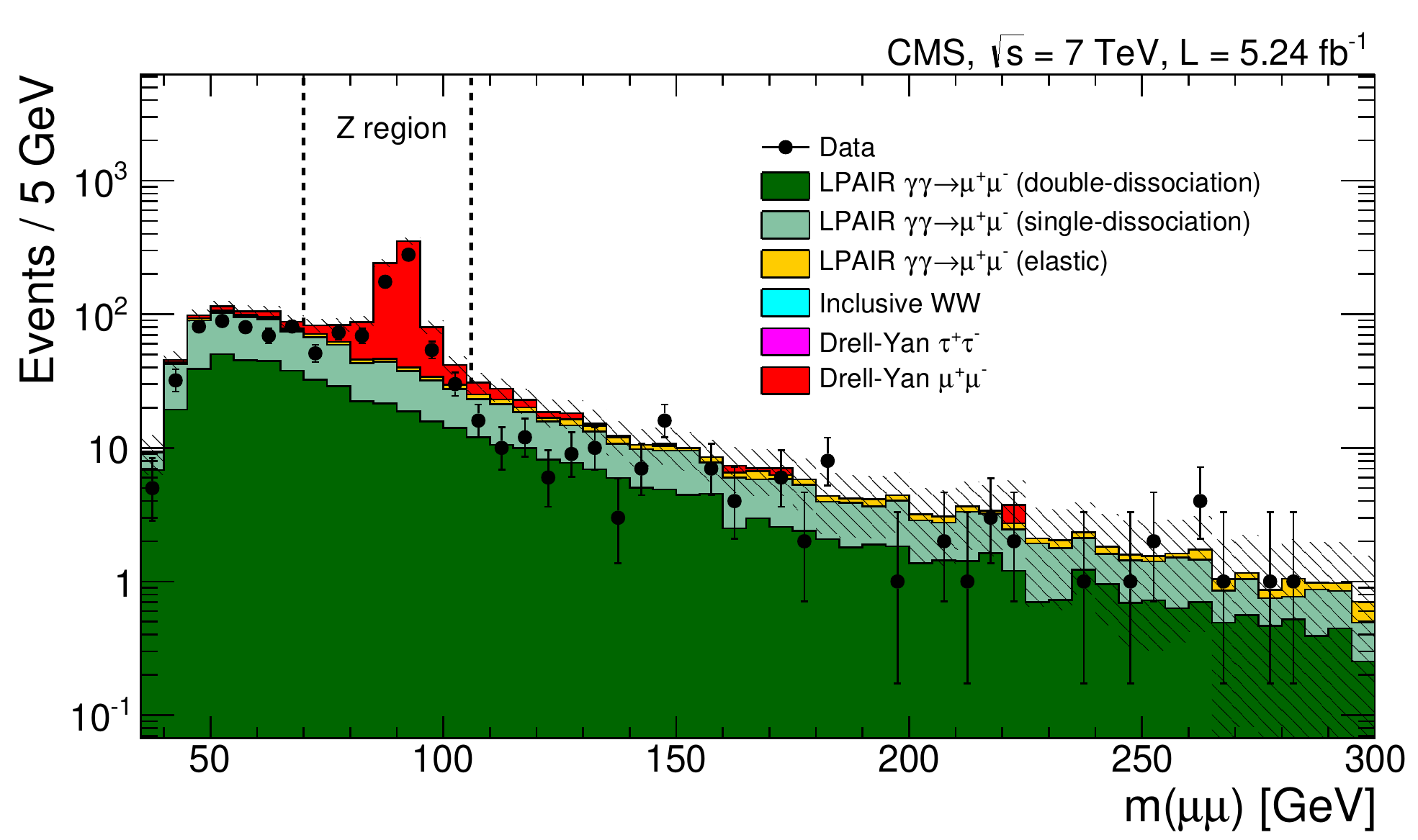}
\caption{\small{Invariant mass distribution of the muon pairs for the dissociation selection. The dashed lines indicate the $\cPZ$-peak region. The hatched bands
indicate the statistical uncertainty in the simulation.}}
\label{fig:mumu-alumi-invmass}
\end{figure}

\begin{figure}[h!t]
\centering
\includegraphics[width=6.5cm]{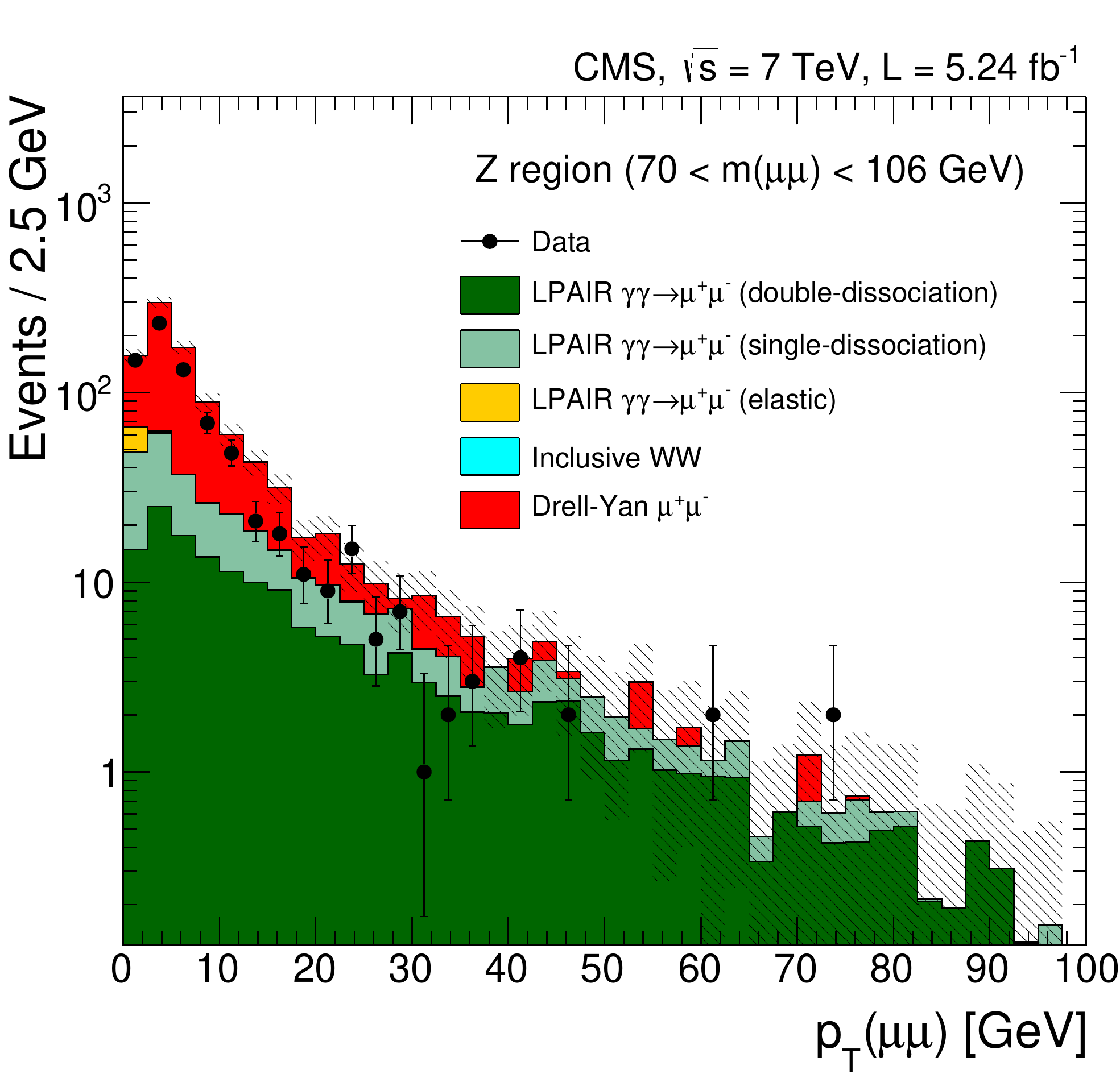}
\includegraphics[width=6.5cm]{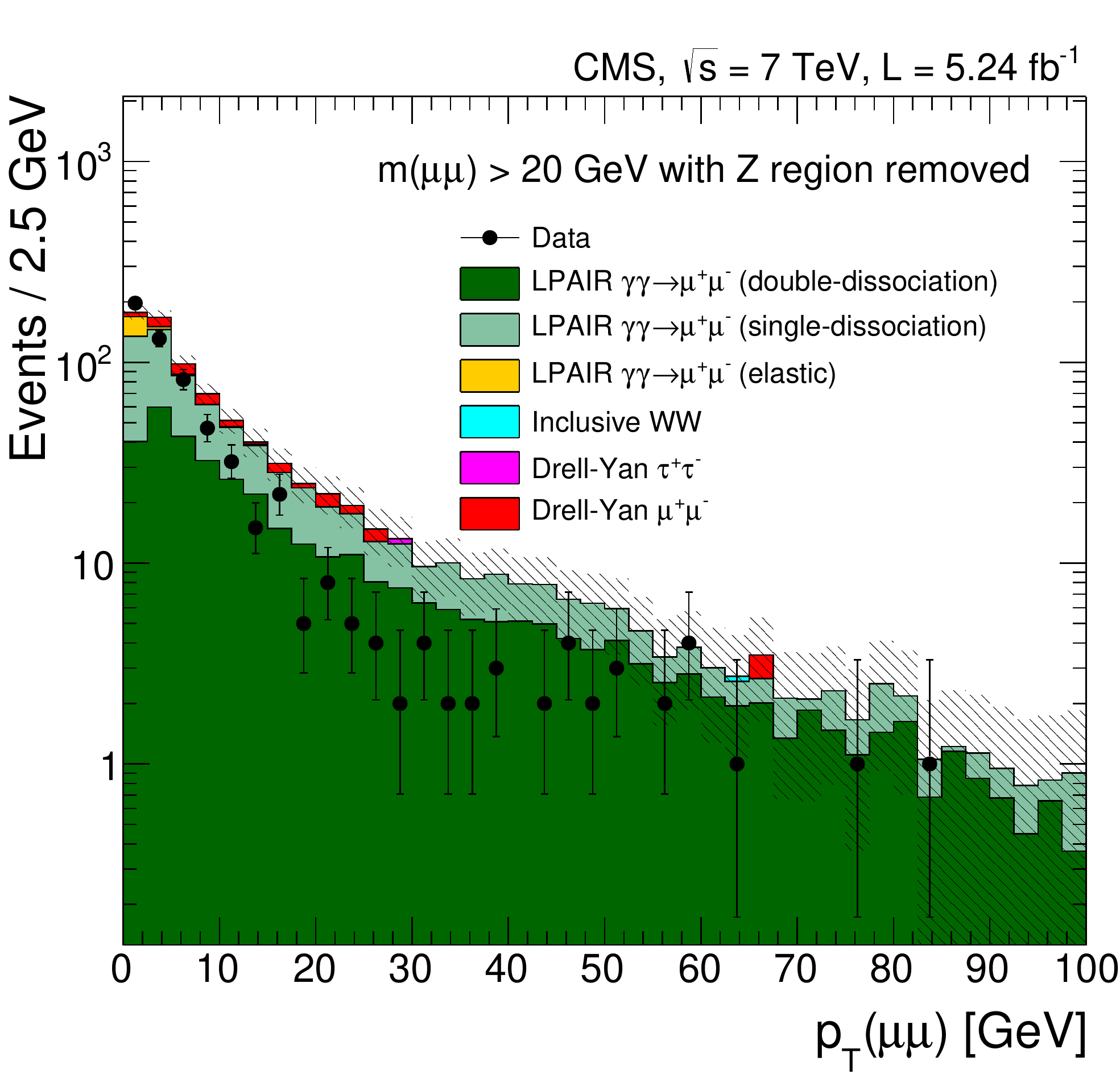}
\caption{\small{Transverse momentum distribution for $\Pgmp\Pgmm$ pairs with zero extra tracks passing the dissociation selection, for
the $\cPZ$ region only (left), and with the $\cPZ$ region removed (right).
The hatched bands indicate the statistical uncertainty in the simulation.}}
\label{fig:mumu-alumi-acopl}
\end{figure}

Table~\ref{tab:mumu-numevts} lists the number of events with zero extra tracks seen in the data and expected
from simulation in the $\Pgmp\Pgmm$ sample after trigger and preselection criteria. In the elastic region, the sum of all contributions in simulation using
the {\textsc{lpair}} generator is $\sim$10\% greater than the yield observed in data. In the dissociation
region, which is expected to be most affected by rescattering corrections~\cite{HarlandLang:2012qz}, an overall deficit of 28\% is observed in the data.
As seen in Fig.~\ref{fig:mumu-alumi-acopl}, this deficit is particularly large at high $\pt(\Pgmp\Pgmm)$.

\begin{table}[thb]
\topcaption{\label{tab:mumu-numevts}Total number of data events compared to the sum of all the background events expected in the two control regions,
after trigger and preselection criteria. The uncertainties are statistical only.}
\centering
\begin{tabular}{l rrrr}
\hline
Region & Data & Simulation & Data/Simulation\\
\hline
Elastic                  & $820$     & $906  \pm 9$     & $0.91 \pm 0.03$\\
Dissociation             & $1312$    & $1830 \pm 17$     & $0.72 \pm 0.02$\\
Total                    & $2132$    & $2736 \pm 19$     & $0.78 \pm 0.02$\\
\hline
\end{tabular}
\end{table}

The suppression due to rescattering corrections is particularly significant in the case of quasi-exclusive production when one or both incident protons
dissociate. This suppression is practically impossible to calculate from first principles as it involves very soft interactions and only phenomenological
models are available. Therefore, using the low-background sample of dimuons produced via two-photon interactions, we use the data to determine an effective,
observed ``luminosity'' of two-photon interactions at high energies relevant for $\PW$-pair production. For this purpose, the number of detected dimuon events
with invariant mass over 160\GeV, corrected for the DY contribution, is divided by the prediction
for the fully exclusive, elastic production predicted by {\textsc{lpair}},
\begin{equation}
\begin{aligned}
F &= \left.\frac{N_{\mu\mu~\text{data}}-N_\mathrm{DY}}{N_\text{elastic}}\right|_{m(\Pgmp\Pgmm)>160\GeV}.\\
F &= 3.23 \pm 0.53.
\label{eq2}
\end{aligned}
\end{equation}

This factor $F$ is then be applied to scale the {\CALCHEP} signal prediction and obtain the total cross section for two-photon $\PWp\PWm$
production including elastic and proton dissociative contributions. This assumes the dilepton kinematics are the same in elastic and
proton dissociative production, with the difference in efficiency arising from the requirement of zero extra tracks
originating from the $\PWp\PWm$ production vertex.

A total uncertainty of 16\% on this factor $F$ is assigned, which has two independent sources. The first source is a 15.5\% statistical uncertainty in
the determination of this factor from the high-mass dimuon data. The second source is due to applying the scale factor derived from the
matrix-element {\textsc{lpair}} generator to the $\gamma\gamma\rightarrow \PWp\PWm$ signal sample produced with {\CALCHEP} according to the
equivalent photon approximation (EPA) \cite{1975PhR....15..181B}. This is checked by comparing the {\textsc{lpair}} prediction with the EPA prediction for muon pair
production above 160\GeV in invariant mass, and is taken conservatively as 5\%.

\section{The \texorpdfstring{$\PWp\PWm \rightarrow \mu^{\pm}\Pe^{\mp}$}{W(+)W(-) to mu(+)mu(-)} signal}
\label{mue}

The SM cross section for the purely elastic process $\Pp\Pp\to \Pp \PWp\PWm \Pp$
is predicted to be 40.0\unit{fb} using {\CALCHEP}, or 1.2\unit{fb} for the cross section times branching fraction to $\mu^{\pm}\Pe^{\mp}$
final states~\cite{Beringer:1900zz}. Using the scale factor $F$ extracted from the high-mass $\gamma\gamma\rightarrow\Pgmp\Pgmm$ sample to account for
the additional proton dissociation contribution, the total predicted cross section times branching fraction is:
\begin{equation*}
\sigma_{\mathrm{theory}}(\Pp\Pp \rightarrow\Pp^{(*)}\PWp\PWm\Pp^{(*)} \rightarrow\Pp^{(*)}\mu^{\pm}\Pe^{\mp}\Pp^{(*)}) = 4.0\pm0.7\unit{fb}.
\end{equation*}
The acceptance for the SM signal in the fiducial region $|\eta (\mu, \Pe)|<2.4$, $\pt(\mu, \Pe)>20\GeV$ is determined to be
55\% using the {\CALCHEP} generator.

\begin{table}
\topcaption{\label{tab:cutflowtable}Product of the signal efficiency and the acceptance, visible cross section, and number of events selected in
data at each stage of the selection. The preselection requires a reconstructed muon and electron of opposite charge, each having $\pt>20\GeV$
and $\abs{\eta}<2.4$, matched to a common primary vertex with fewer than 15 additional tracks.}
\centering
\begin{tabular}{l c c c}
  \hline
  Selection step & Signal $\epsilon \times A$ & Visible cross section (fb) & Events in data \\
  \hline
 Trigger and preselection                         & 28.5\%                    & 1.1                        & 9086           \\
 $m(\mu^{\pm}\Pe^{\mp})> 20\GeV$                    & 28.0\%                    & 1.1                        & 8200           \\
 Muon ID and Electron ID                          & 22.6\%                    & 0.9                        & 1222           \\
 $\mu^{\pm}\Pe^{\mp}$ vertex with zero extra tracks & 13.7\%                    & 0.6                        & 6              \\
 $\pt(\mu^{\pm}\Pe^{\mp})>30\GeV$                  & 10.6\%                    & 0.4                        & 2              \\
  \hline
\end{tabular}
\end{table}

The predicted visible cross section at each stage of the selection, defined as the predicted cross section multiplied by the efficiency and acceptance, is
shown in Table~\ref{tab:cutflowtable}, together with the efficiency and acceptance for the signal, and the corresponding number of events selected from
the data sample. The signal inefficiency introduced by the requirement of zero extra tracks on the $\mu^{\pm}\Pe^{\mp}$ vertex reflects the effect of pileup.
As described in Ref.~\cite{Chatrchyan:2011ci}, with increasing pileup there is a higher probability of finding tracks from a pileup interaction in close
proximity to the dilepton signal vertex. The incorrect assignment of these tracks to the signal vertex by the vertex clustering algorithm will lead to
signal events being rejected. In the 2011 data sample with an average of 9 interactions per bunch crossing, this results in the rejection of
$\sim$40\% of signal events that would pass other selection requirements.

To check the modelling of the individual background contributions, we define three independent control regions based on the number of tracks
associated to the $\mu^{\pm}\Pe^{\mp}$ vertex and the \pt of the $\mu^{\pm}\Pe^{\mp}$ pair, as defined in Table~\ref{controlregiondefinitions}. To
study the inclusive backgrounds, we select two control regions with 1--6 extra tracks associated to the $\mu^{\pm}\Pe^{\mp}$ vertex. The first
region has $\pt(\mu^{\pm}\Pe^{\mp})<30\GeV$ and is dominated by inclusive Drell--Yan production of $\tau^{+}\tau^{-}$ and the second, with
$\pt(\mu^{\pm}\Pe^{\mp})>30\GeV$, is dominated by inclusive $\PWp\PWm$ production. In order to select a sample with a significant fraction
of $\gamma\gamma\rightarrow\tau^{+}\tau^{-}$ events, we define a third control region having zero extra tracks, but $\pt(\mu^{\pm}\Pe^{\mp})<30\GeV$.

\begin{table}
\topcaption{\label{controlregiondefinitions}Definitions for the three independent control regions.}
\centering
\begin{tabular}{c c c c}
  \hline
  Region & Background process & $N_{\mathrm{extra~tracks}}$ & $\pt(\mu^{\pm}\Pe^{\mp})$ \\
  \hline
  1      & Inclusive $\PWp\PWm$                             & $1 \leq N_{\text{extra~tracks}} \leq 6$ & $>$30\GeV \\
  2      & Inclusive Drell--Yan $\tau^+\tau^-$             & $1 \leq N_{\text{extra~tracks}} \leq 6$ & $<$30\GeV \\
  3      & $\gamma\gamma\to\tau^+\tau^-$                  & $N_{\text{extra~tracks}} = 0$           & $<$30\GeV \\
  \hline
\end{tabular}
\end{table}

We first compare the data to the expected backgrounds from simulation in the inclusive $\PWp\PWm$ region.
The predicted \textsc{pompyt} diffractive $\PWp\PWm$ contribution is, very conservatively, added to the other backgrounds,
without accounting for any survival probabilities or overlap with the inclusive $\PWp\PWm$ sample. To study
the $\PW+\text{jets}$ backgrounds, for which the contribution is mainly from misidentified leptons or non-prompt leptons in jets, we select a
control sample of events with $\pt(\mu^{\pm}\Pe^{\mp})>30\GeV$, where at least one of the two lepton candidates fails the nominal
offline identification criteria. This sample is then normalized to the simulation in the high-multiplicity (more than 6 extra tracks) region
and used to estimate the $\PW$+jets background in the signal and inclusive $\PWp\PWm$ control regions. Figure~\ref{fig:nextratrackspt30} shows
the distribution of the number of extra tracks for the $\PWp\PWm$ region with $\pt(\mu^{\pm}\Pe^{\mp})>30\GeV$, together with
the invariant mass and acoplanarity of the events with 1--6 extra tracks. In general the data are consistent with the sum of
simulated backgrounds in this region.

\begin{figure}[h!t]
\centering
\includegraphics[width=6.5cm]{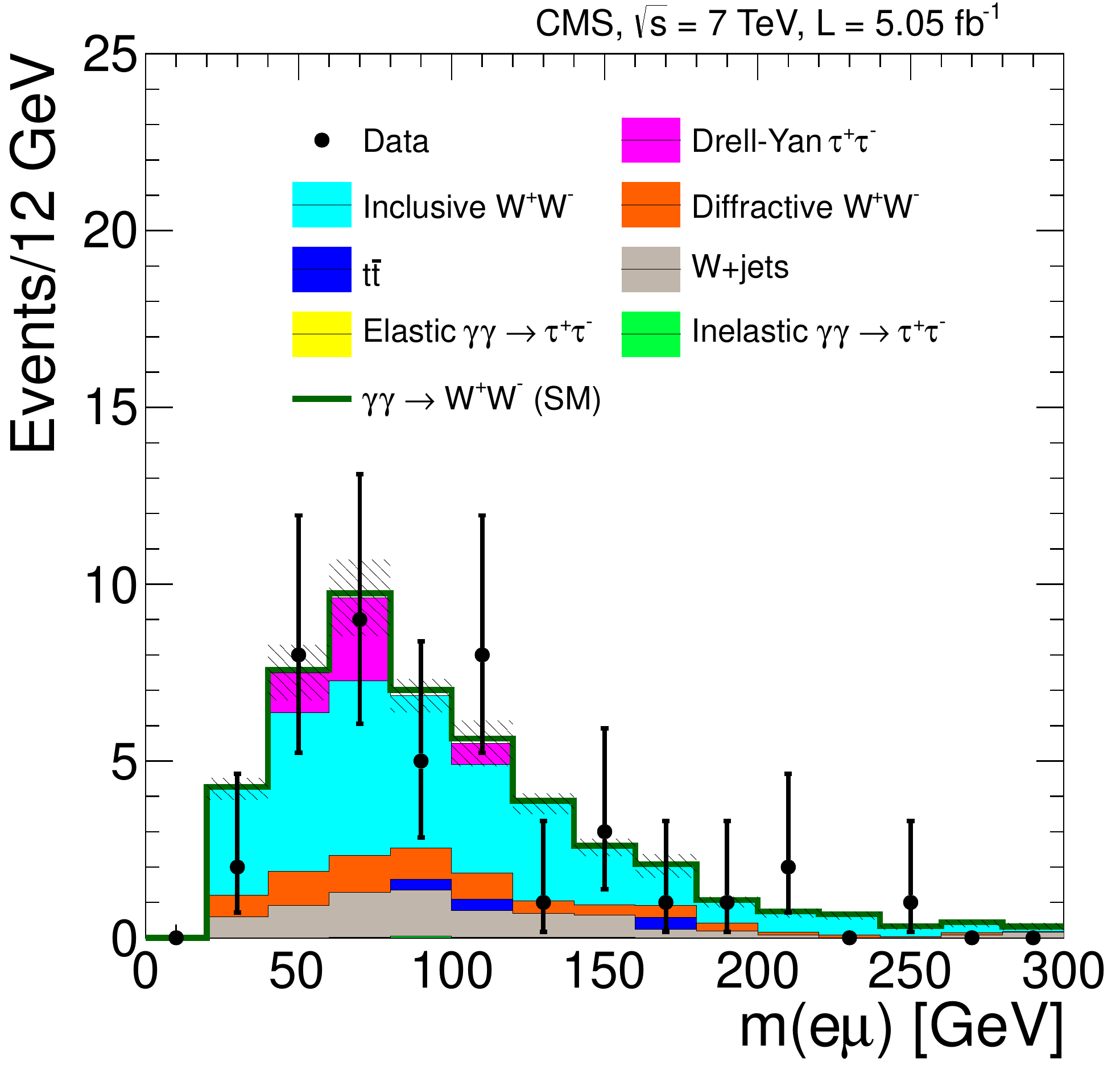}
\includegraphics[width=6.5cm]{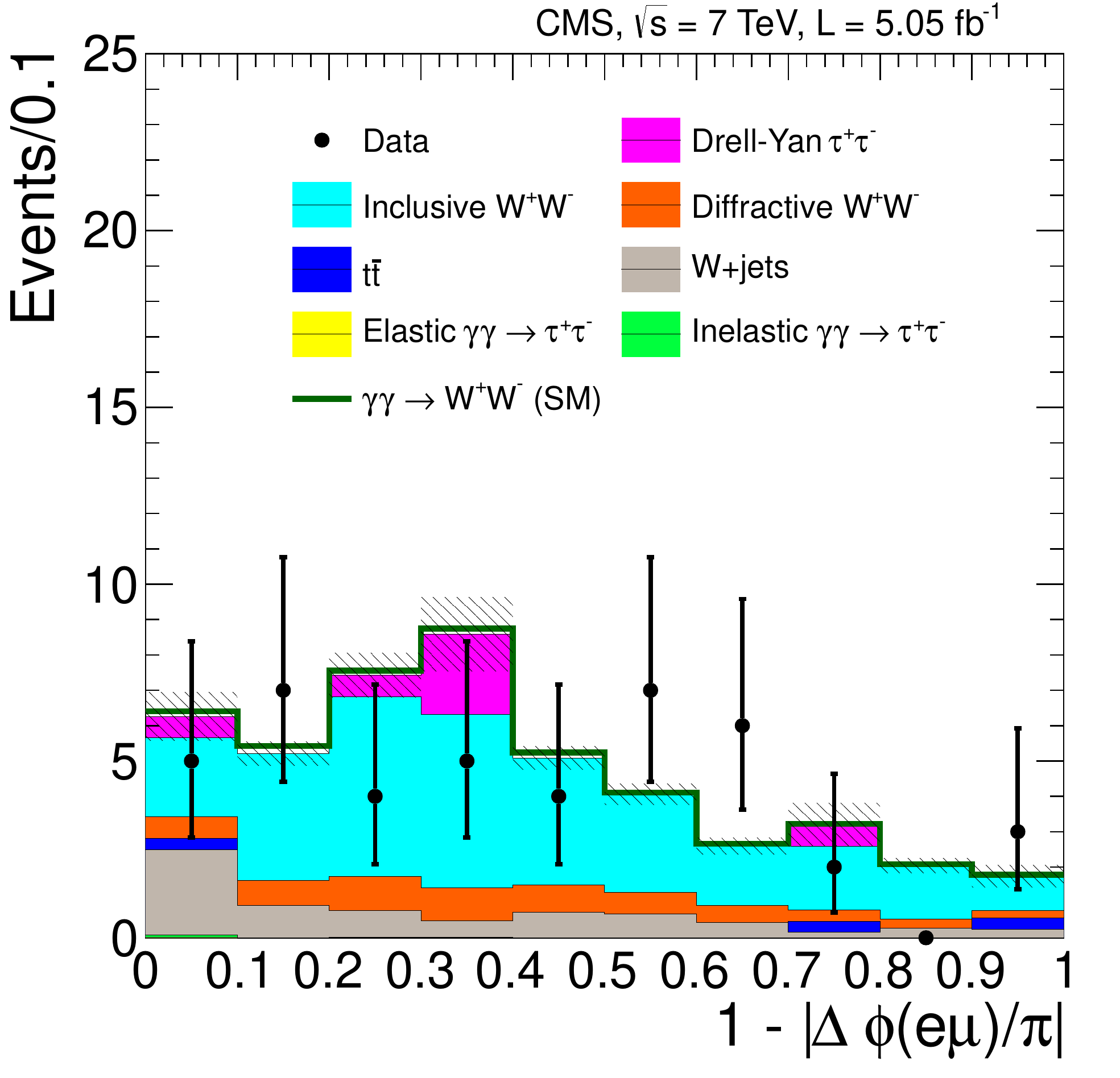}
\includegraphics[width=6.5cm]{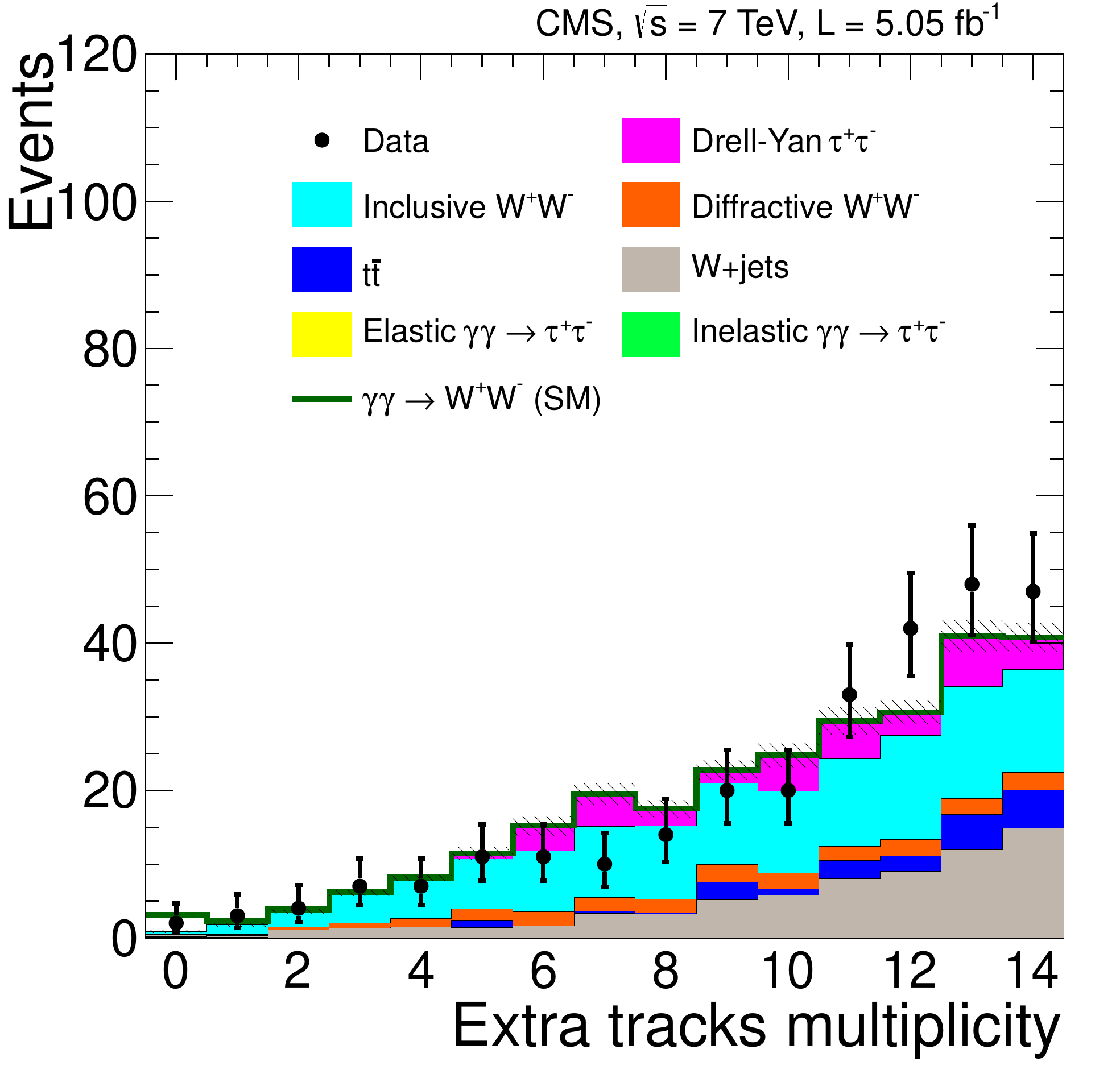}
\caption{\small{Data compared to simulation in control region 1. The $\mu^{\pm}\Pe^{\mp}$ invariant mass (above left) and acoplanarity (above right)
are shown for events with 1--6 extra tracks on the $\mu^{\pm}\Pe^{\mp}$ vertex and $\pt(\mu^{\pm}\Pe^{\mp})>30\GeV$. The number of additional tracks on
the electron-muon primary vertex is shown for events with $\pt(\mu^{\pm}\Pe^{\mp})>30\GeV$ (below). The shaded bands indicate the statistical uncertainty
in the background estimation. The signal (open histogram) is shown stacked on top of the backgrounds.}}
\label{fig:nextratrackspt30}
\end{figure}

In the Drell--Yan $\tau^{+}\tau^{-}$-dominated region with $\pt(\mu^{\pm}\Pe^{\mp})<30\GeV$ and 1--6 tracks we find general agreement
in the dilepton kinematic distributions, but an overall deficit in the data sample compared to simulation, with $256.7 \pm 10.1$ background
events expected and 182 observed. Figure~\ref{fig:nextratracksptlessthan30} shows the distribution of the number of extra tracks for the events with
with $\pt(\mu^{\pm}\Pe^{\mp})<30\GeV$, together with the invariant mass and acoplanarity of the events with 1--6 extra tracks.

\begin{figure}[h!t]
\centering
\includegraphics[width=6.5cm]{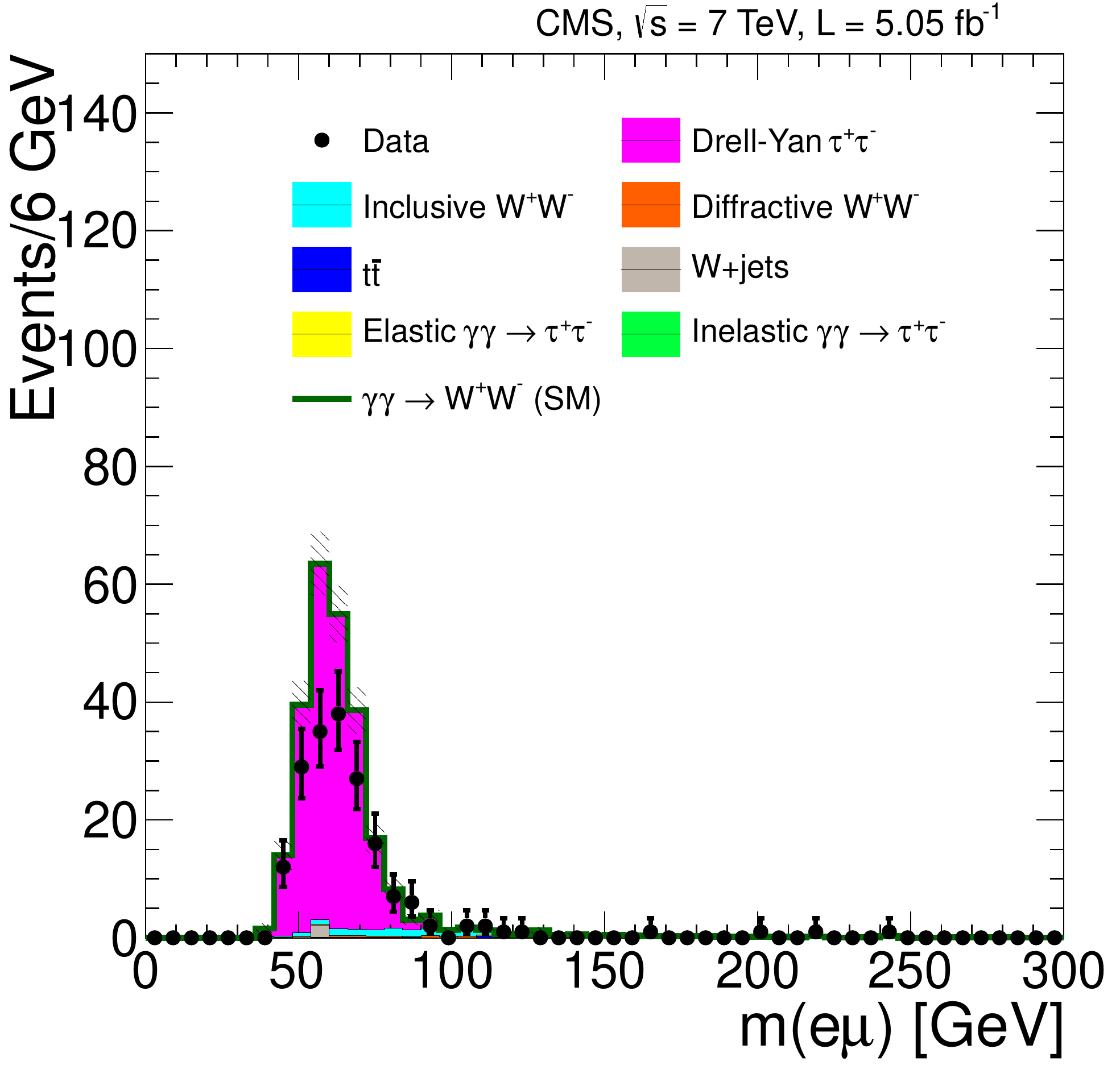}
\includegraphics[width=6.5cm]{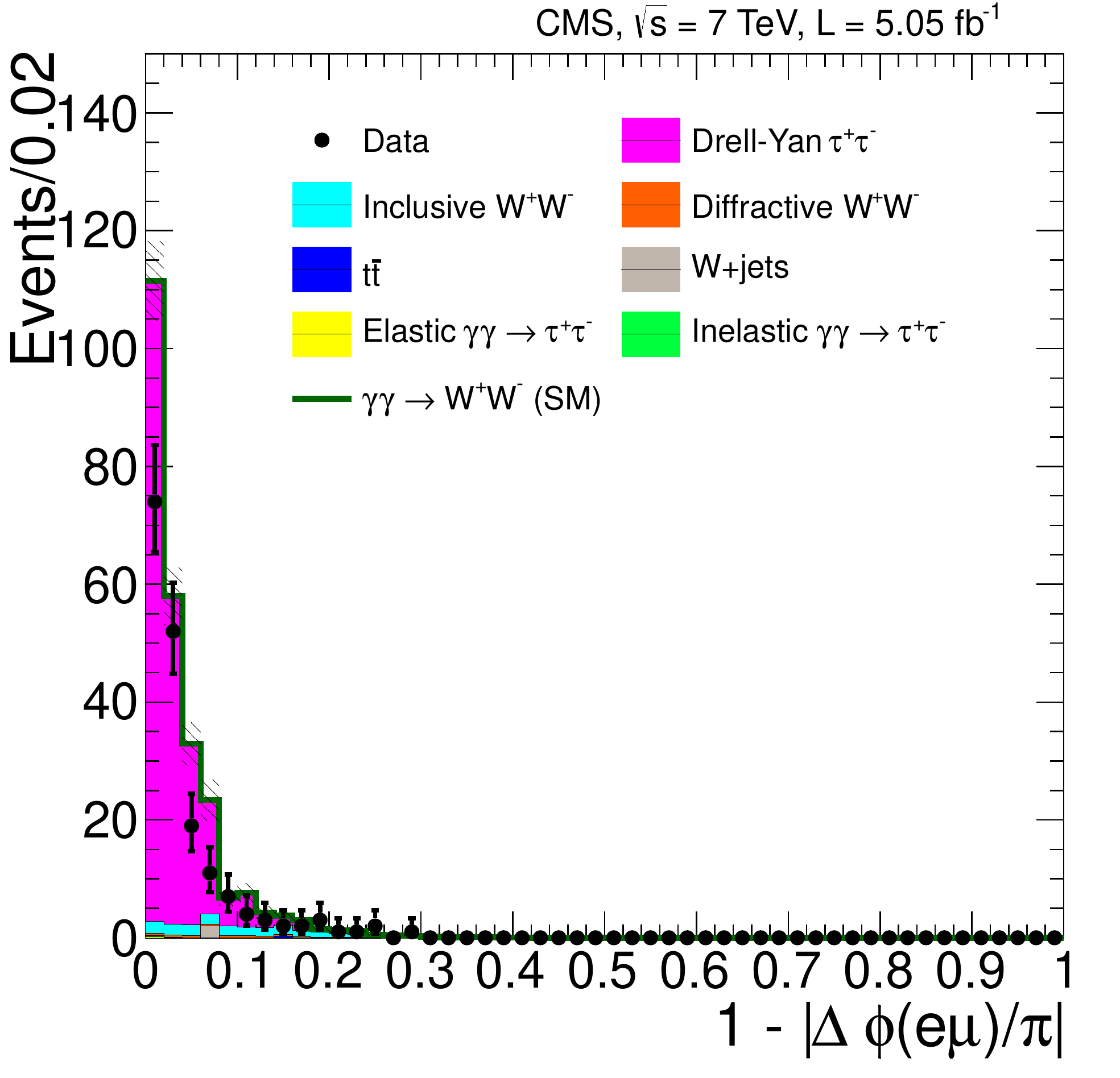}
\includegraphics[width=6.5cm]{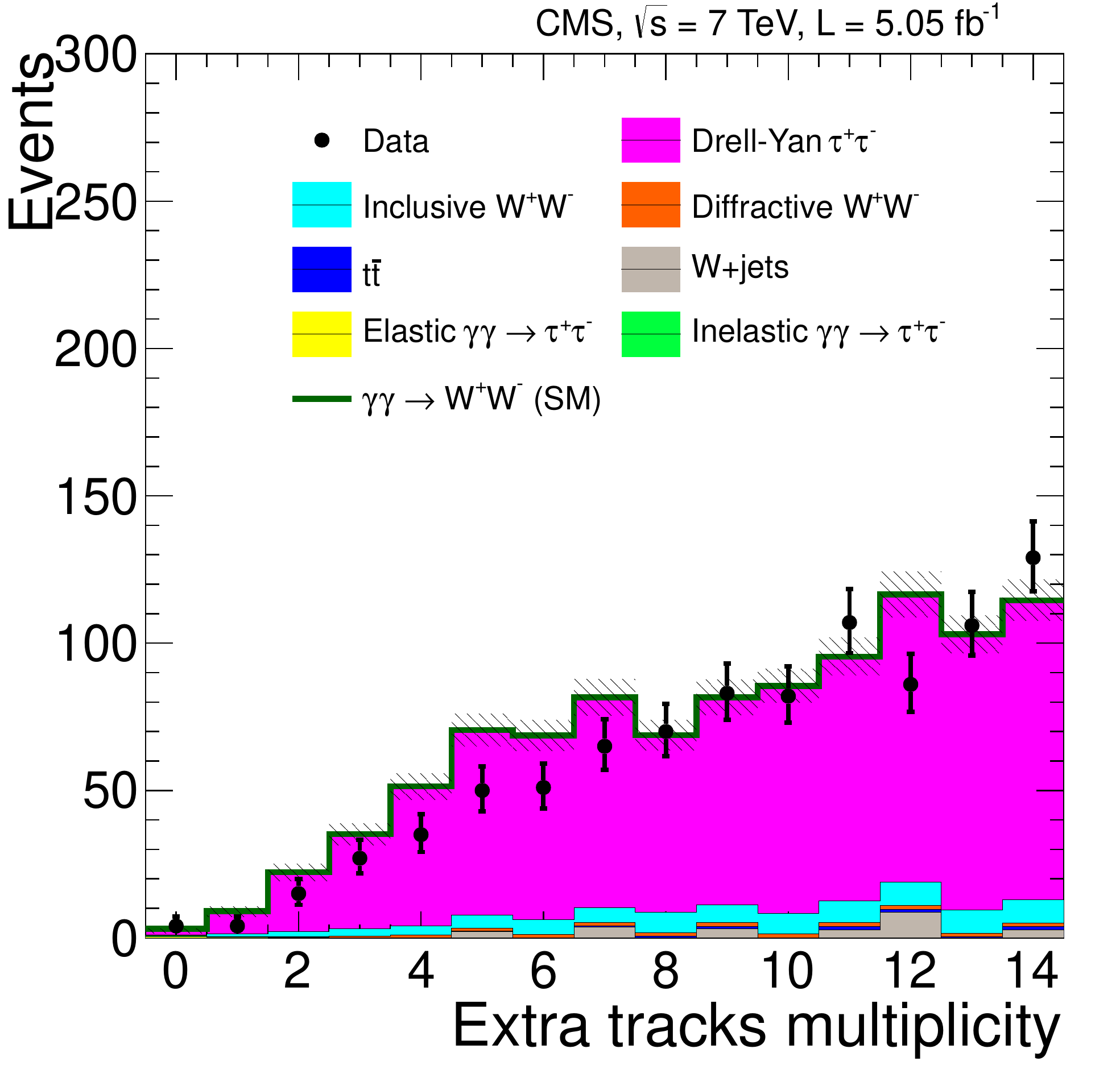}
\caption{\small{Data compared to simulation in control region 2. The $\mu^{\pm}\Pe^{\mp}$ invariant mass (above left) and acoplanarity (above right)
are shown for events with 1--6 extra tracks on the $\mu^{\pm}\Pe^{\mp}$ vertex and $\pt(\mu^{\pm}\Pe^{\mp})<30\GeV$. The number of additional tracks on
the electron-muon primary vertex is shown for events with $\pt(\mu^{\pm}\Pe^{\mp})<30\GeV$ (below). The shaded bands indicate the statistical uncertainty
in the background estimation. The signal (open histogram) is shown stacked on top of the backgrounds.}}
\label{fig:nextratracksptlessthan30}
\end{figure}

In the $\tau^{+}\tau^{-}$ sample with zero extra tracks, we find four events in the data sample, compared to a background expectation of 2.5~events from simulation,
plus 0.9 events from the $\gamma\gamma\rightarrow \PWp\PWm$ signal. The expected contribution to the background from
$\gamma\gamma\rightarrow\tau^{+}\tau^{-}$ is approximately 0.7 events. The invariant mass and acoplanarity distributions are shown in
Fig.~\ref{fig:memudphipt0trackslessthan30}.

Table~\ref{controlregionyields} summarizes the observed and expected background event yields for the three
independent control regions. Tracks from pileup vertices may be wrongly associated to the $\mu^{\pm}\Pe^{\mp}$
vertex from a $\gamma\gamma\rightarrow~\PWp\PWm$ event, resulting in signal events being classified as 1--6 tracks events.
This signal contamination, as well as that from signal events with $\pt(\mu^{\pm}\Pe^{\mp})<30\GeV$,
is estimated from simulation to be approximately one event or less in any of the control regions.

\begin{figure}[h!t]
\centering
\includegraphics[width=6.5cm]{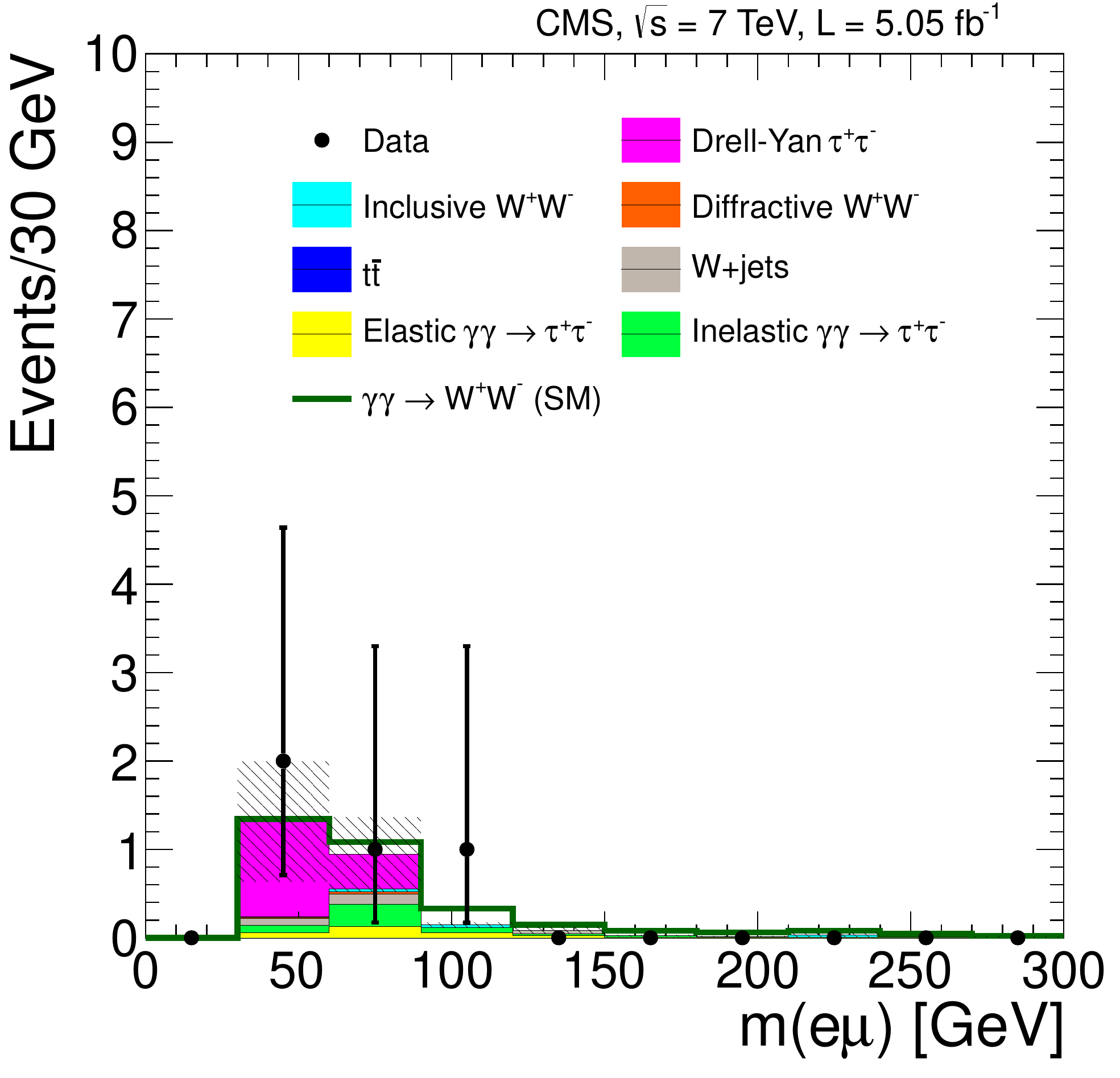}
\includegraphics[width=6.5cm]{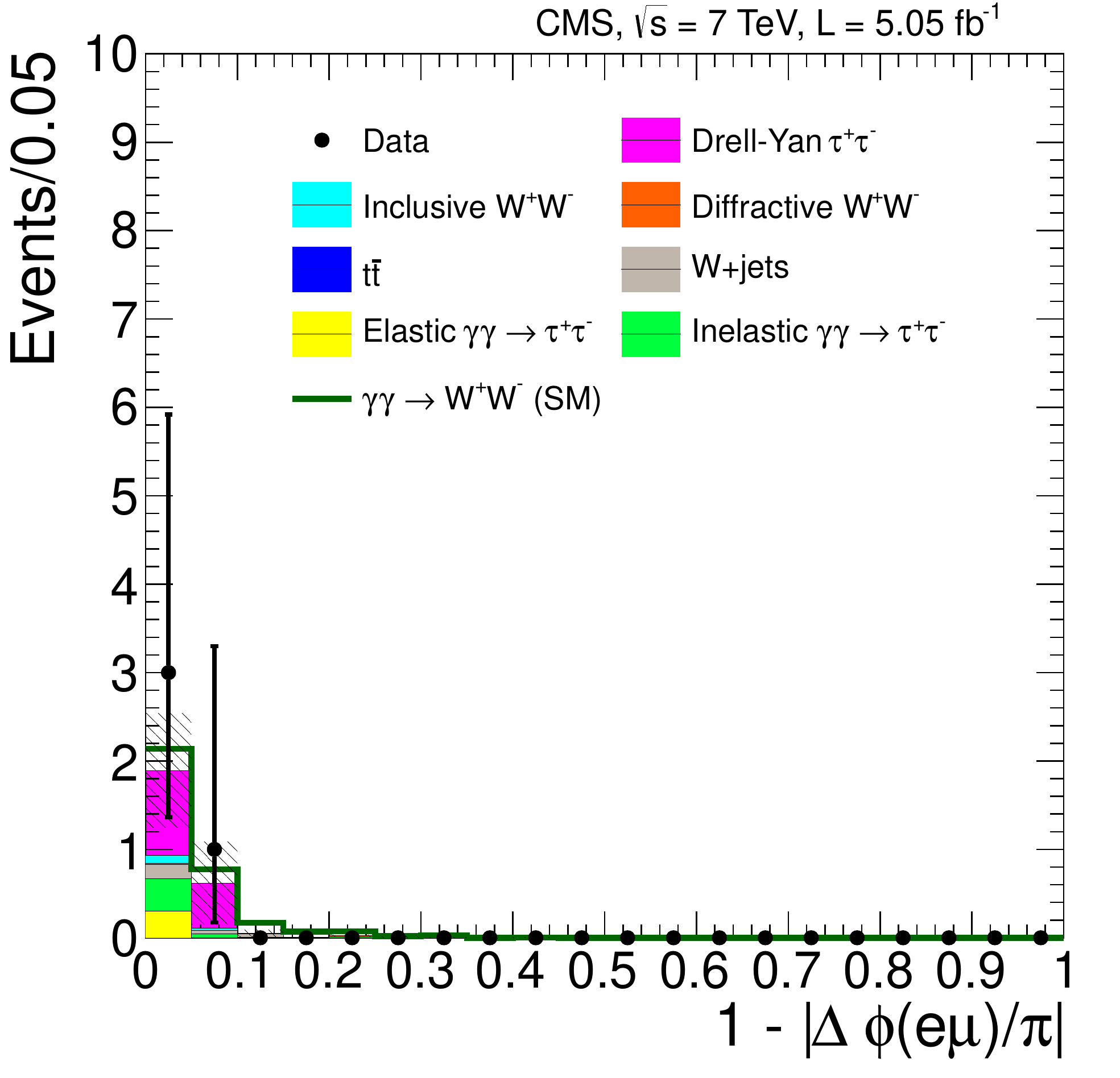}
\caption{\small{Data compared to simulation for control region 3. The $\mu^{\pm}\Pe^{\mp}$ invariant mass (left) and acoplanarity (right) are displayed
for events with zero extra tracks on the $\mu^{\pm}\Pe^{\mp}$ vertex and $\pt(\mu^{\pm}\Pe^{\mp})<30\GeV$. The shaded bands indicate the
statistical uncertainty in the background estimation. The signal (open histogram) is shown stacked on top of the backgrounds.}}
\label{fig:memudphipt0trackslessthan30}
\end{figure}

\begin{table}
\topcaption{\label{controlregionyields}Background event yields for the three independent control regions.}
\centering
\begin{tabular}{c c c c c}
  \hline
  Region & Background process & Data & Sum of backgrounds & $\gamma\gamma\rightarrow \PWp\PWm$ signal \\
  \hline
  1 & Inclusive $\PW^+W^-$                      & 43   & $46.2  \pm 1.7$       & 1.0           \\
  2 & Inclusive Drell--Yan $\tau^+\tau^-$      & 182  & $256.7 \pm 10.1$      & 0.3           \\
  3 & $\gamma\gamma\to\tau^+\tau^-$           & 4    & $2.6   \pm 0.8$       & 0.7           \\
  \hline
\end{tabular}
\end{table}

We use the simulated background sample, corrected for trigger and lepton identification efficiencies, to estimate the backgrounds
in the signal region. The $\PW+\text{jets}$ contribution to the background is estimated from the control sample of events with lepton identification
inverted, while the $\gamma\gamma\rightarrow\tau^{+}\tau^{-}$ contribution is normalized using the factor derived from the high-mass
$\gamma\gamma\rightarrow\Pgmp\Pgmm$ data sample. No simulated Drell--Yan $\tau^+\tau^-$ events survive all selection criteria,
and a deficit of data compared to simulation is observed in the corresponding control region with 1--6 extra tracks and $\pt(\mu^{\pm}\Pe^{\mp})<30\GeV$.
Given this, and the agreement with data in the $\PWp\PWm$ control region with 1--6 extra tracks and the $\tau^{+}\tau^{-}$ region with zero extra
tracks, no additional rescaling of the backgrounds is performed. The estimated background is $0.84\pm0.15$ events, including the systematic
uncertainty on the backgrounds.

\section{Systematics and cross-checks}
\label{systematics}

The systematic uncertainties affecting the signal are summarized in Table~\ref{tab:signalsyst}. The uncertainty on the delivered 2011 luminosity is
2.2\%~\cite{CMSPixelLumi}. The lepton trigger and selection efficiency corrections are varied by their $\pm 1\sigma$ statistical uncertainties, with the direction
of the variation within each $\pt$ and $\eta$ bin correlated.
The largest variation in the expected signal (when varying the efficiency scale factors by ${+}1\sigma$) is 4.2\%, which is taken as a systematic uncertainty on the signal yield.
The variation in the sum of backgrounds expected from simulation due to the trigger and lepton selection is 3.7\%, which is taken as a component of the systematic uncertainty on the background estimate. The uncertainty on the efficiency for reconstructing vertices with two tracks is estimated to be 1.0\%, based on the data
vs. simulation difference obtained from the method described in Ref.~\cite{CMS:2010wta}.

The efficiency of the exclusivity selection, including effects from pileup, is checked using the $\gamma\gamma\rightarrow\Pgmp\Pgmm$ control sample. Using the
elastic control region, where the theoretical uncertainties are smallest, we assign a 10\% systematic uncertainty based on the level of agreement between data and
simulation. In addition, we check the stability of the agreement between data and simulation as a function of pileup, using samples ranging from a minimum of
1-5 reconstructed vertices to a maximum of 11--20.

The predictions for both the $\gamma\gamma\rightarrow \PWp\PWm$ signal and the $\gamma\gamma\rightarrow\tau^{+}\tau^{-}$
background are rescaled to reflect the contribution of proton dissociation, as derived from the high-mass $\gamma\gamma\rightarrow\Pgmp\Pgmm$ sample.
As described in Section~\ref{sect:mumu-splitted-runepochs}, a total uncertainty of 16\% is assigned to this factor scale factor $F$, based on the
statistical uncertainty of the high-mass $\gamma\gamma\rightarrow\Pgmp\Pgmm$ control sample and the difference between the matrix-element and EPA
approaches.

As a cross-check we perform several alternative estimates and tests of the nominal background contribution of $0.84\pm0.15$ events. To
check the sensitivity to the simulation of the dominant $\PWp\PWm$ background, we replace the default \MADGRAPH sample with
a \PYTHIA sample normalized to the NLO cross section. The agreement with data in the control region is similar to that of {\MADGRAPH}
and results in a total background estimate of $0.71\pm0.21\stat$ events in the signal region. Scaling the inclusive
$\PWp\PWm$ background to the central value of the CMS cross section measurement~\cite{Chatrchyan:2013yaa}, rather than the NLO prediction, would change
the total background estimate to $0.88\pm0.15$ events. This change is smaller than the uncertainty on the nominal estimate.
The sensitivity to the diffractive component of the $\PWp\PWm$ background is further tested by varying the cross section
between 0\% and 200\% of the nominal value. This results in a variation of $\pm$0.03~events in the total background estimate.
The contribution from vector boson fusion (VBF), $\PW\PW \rightarrow \PW\PW$, is estimated using the \textsc{vbfnlo} event generator.
No VBF events survive all selections, corresponding to an upper limit of $\sim$0.1 events at 95\% Confidence Level (CL). The contribution of VBF
in the 1--6 tracks control region is estimated to be approximately 0.1 events.

In addition, we take advantage of the lack of correlation between the number of the extra tracks and $\pt(\mu^{\pm}\Pe^{\mp})$ in the main background
processes to estimate the background from data using the three control regions defined in Section~\ref{mue}. With uncorrelated variables
the relationship between the number of events in each region can be expressed as $N_{D}/N_{A} = N_{B}/N_{C}$, where $N_{A}$, $N_{B}$, and $N_{C}$ represent
the backgrounds in the inclusive $\PWp\PWm$, $\gamma\gamma\to\tau^+\tau^-$, and inclusive Drell--Yan $\tau^+\tau^-$ control regions, respectively, and
$N_{D}$ represents the background in the signal region. The background in the signal region is then obtained by solving for $N_{D}$, resulting in the expression
$N_{D} = (N_{A} \times N_{B})/N_{C}$. After subtracting the signal contamination estimated from simulation in each region, the resulting background estimate
is $0.77\pm0.44\stat$~events, with a large statistical uncertainty due to the low statistics in the $\gamma\gamma\to\tau^+\tau^-$ control region.

We also examine same-sign $\mu^{\pm}\Pe^{\pm}$ events in data to check for possible backgrounds not included in the simulation, because
the main $\PWp\PWm$ and $\tau^{+}\tau^{-}$ backgrounds considered in the analysis are producing opposite-sign
lepton pairs. In the control region which has events containing 1--6 extra tracks we find 8 same-sign events with $\pt(\mu^{\pm}\Pe^{\pm})>30\GeV$ passing all
selection criteria and 11~events with $\pt(\mu^{\pm}\Pe^{\pm})<30\GeV$. No events with fewer than two extra tracks on the
$\mu^{\pm}\Pe^{\pm}$ vertex are observed in the full data sample.

Although no Drell--Yan $\tau^{+}\tau^{-}$ events from the simulation survive all selection criteria, the largest discrepancy between the data sample
and the simulated sample is seen in the corresponding control region with 1--6 extra tracks and $\pt(\mu^{\pm}\Pe^{\mp})<30\GeV$.
Therefore we perform a final check by recalculating the $\tau^{+}\tau^{-}$ backgrounds using an ``embedding'' procedure, in which $\Pgmp\Pgmm$
events are selected in data, and the muons replaced with simulated $\tau$ decays to final states containing an electron and a muon~\cite{Chatrchyan:2012vp}.
In the Drell--Yan control region the embedded sample predicts $165\pm4$ events compared to 182 observed in the data sample. In the signal
region, the total background estimated using the embedded sample is 0.67~$\pm$~0.15 events, which is consistent with the nominal background estimate.

\begin{table}
\topcaption{\label{tab:signalsyst}Summary of systematic uncertainties.}
\centering
\begin{tabular}{lcc}
\hline
                                              & Signal uncertainty & Background uncertainty (events)\\
\hline
Trigger and lepton identification             & 4.2\%  & 0.02  \\
Luminosity                                    & 2.2\%  & 0.005 \\
Vertexing efficiency                          & 1.0\%  & 0.005 \\
Exclusivity and pileup dependence             & 10.0\% & 0.05  \\
Proton dissociation factor                    & 16.3\% & 0.02  \\ \hline
\end{tabular}
\end{table}

The uncertainty in the background estimate includes the statistical uncertainty of the simulated samples or
control samples used to evaluate the backgrounds in the signal region. The uncertainties due to trigger and lepton identification,
vertexing efficiency, and the exclusivity selection are also applied to the backgrounds that are taken from simulation.
An additional uncertainty of $16\%$ is assigned to the $\gamma\gamma\rightarrow\tau^{+}\tau^{-}$ background, reflecting
the uncertainty in the normalization of the proton dissociation contribution derived from the $\gamma\gamma\rightarrow\Pgmp\Pgmm$
control sample.

\section{Results}
\label{results}

Examining the SM $\gamma\gamma\rightarrow \PWp\PWm$ signal region, we find two events passing all the selection criteria, compared to the
expectation of $2.2\pm0.4$ signal events and $0.84\pm0.15$ background events, including the systematic uncertainties listed in
Table~\ref{tab:signalsyst}.

\begin{figure}[h!t]
\centering
\includegraphics[width=6.5cm]{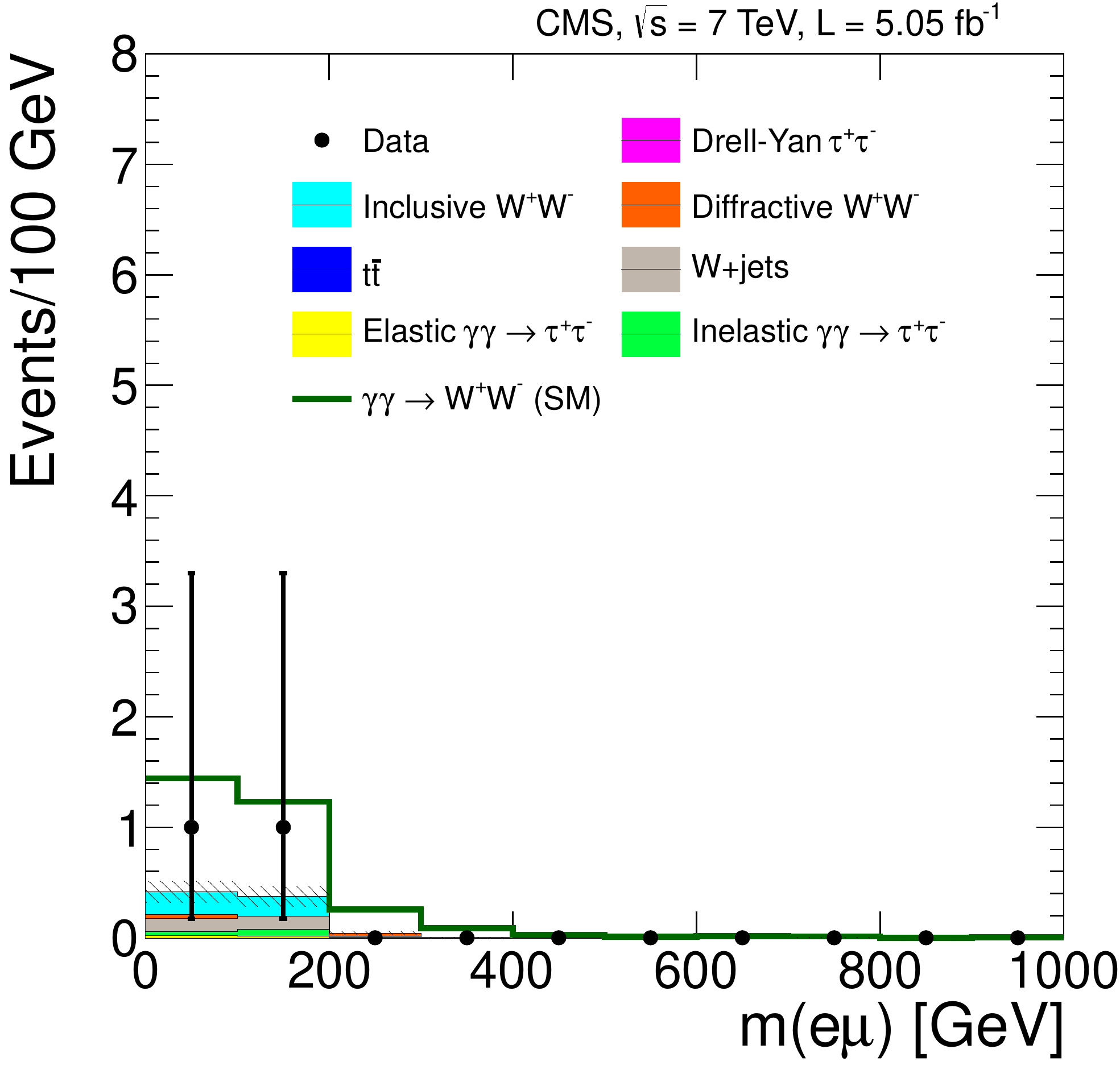}
\includegraphics[width=6.5cm]{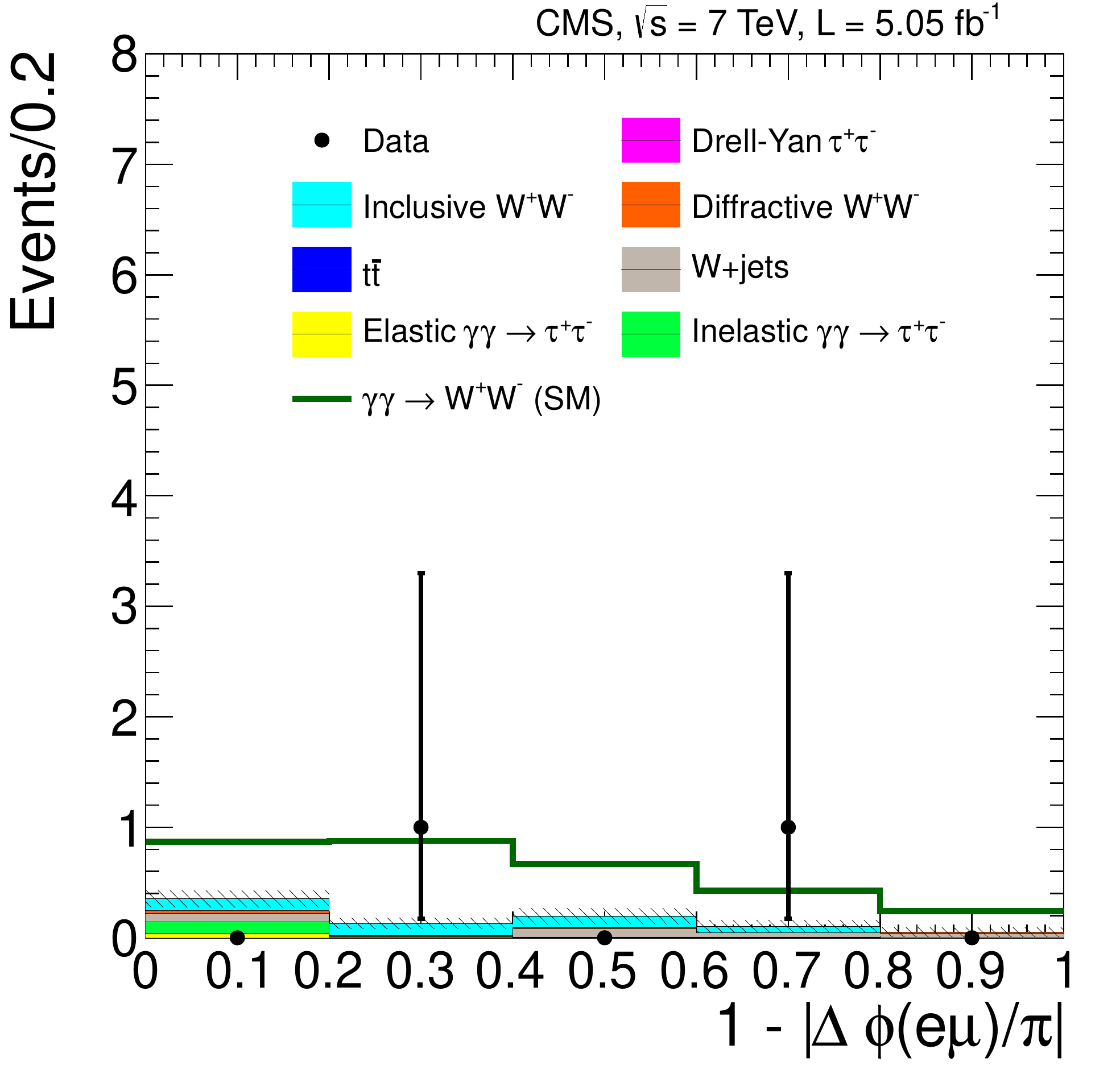}
\includegraphics[width=6.5cm]{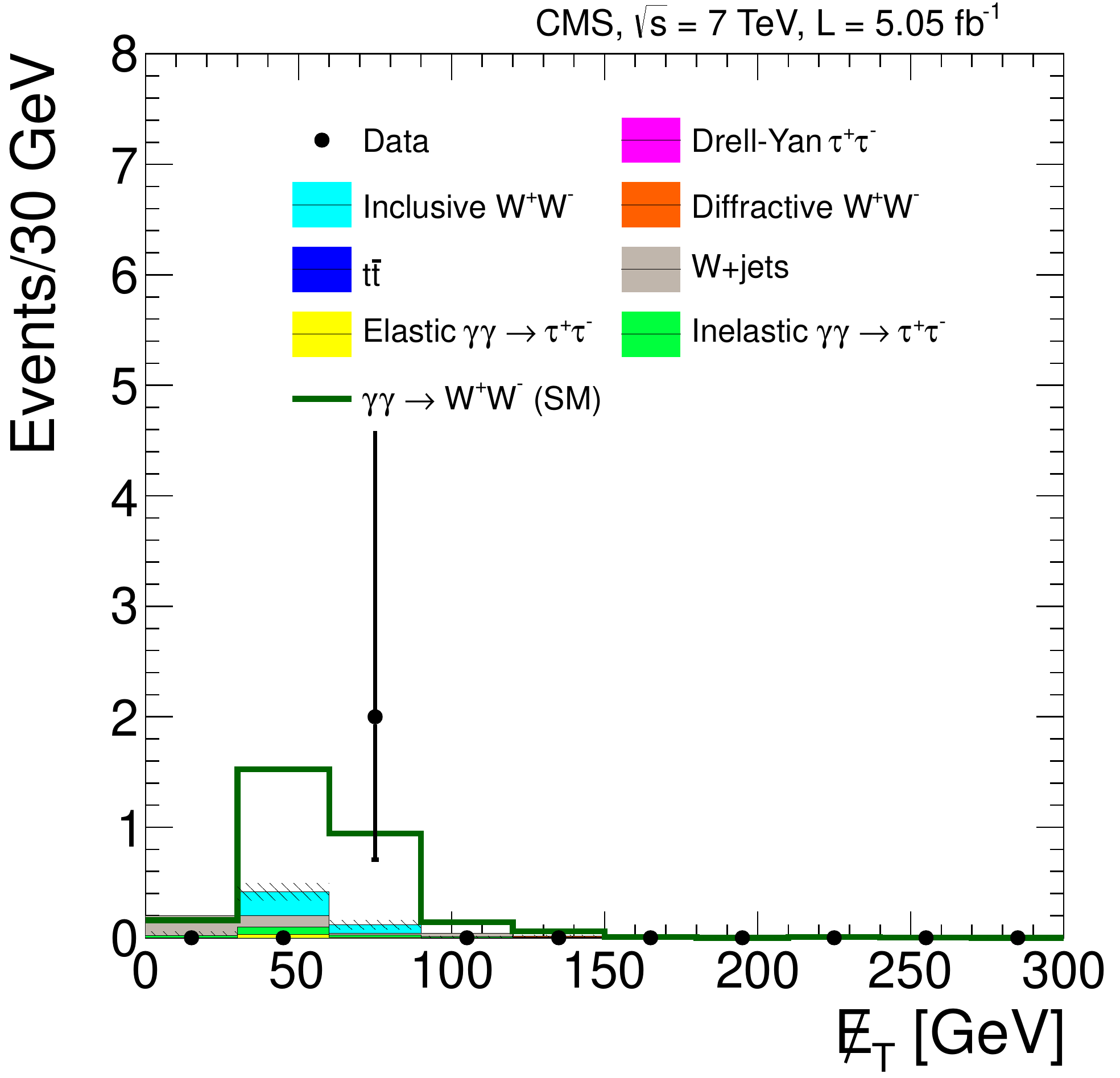}
\caption{\small{The $\mu^{\pm}\Pe^{\mp}$ invariant mass (top left), acoplanarity (top right), and missing transverse energy (bottom) distributions, for events
in the signal region with zero extra tracks on the $\mu^{\pm}\Pe^{\mp}$ vertex and
$\pt(\mu^{\pm}\Pe^{\mp})>30\GeV$. The backgrounds (solid histograms) are stacked with statistical uncertainties indicated by the shaded region, the
signal (open histogram) is stacked on top of the backgrounds.}}
\label{fig:memudphipt0trackspt30}
\end{figure}

We convert the observed results into a cross section and upper limit for events with zero extra tracks
within $\abs{\eta}<2.4$, using the expression $ \sigma = \frac{N}{\epsilon \times A \times \mathcal{L}}$, where $N$ is the number of events observed,
and $\epsilon\times A$ is the efficiency times acceptance for a SM-like signal. Correcting for efficiency, acceptance, and backgrounds, the best
fit signal cross section times branching fraction is:
\begin{equation*}
\sigma (\Pp\Pp \rightarrow\Pp^{(*)}\PWp\PWm\Pp^{(*)} \rightarrow\Pp^{(*)}\mu^{\pm}\Pe^{\mp}\Pp^{(*)}) = 2.2^{+3.3}_{-2.0}\unit{fb},
\end{equation*}
with a significance of ${\sim}1\sigma$. With statistical uncertainties only, the resulting value of the cross section times branching fraction
is $2.2^{+3.2}_{-2.0}\stat\unit{fb}$.

The observed upper limit is estimated using the Feldman--Cousins method~\cite{Feldman:1997qc} to be
2.6 times the expected SM yield at 95\% CL. The median expected limit in the absence of signal
is $1.5^{+1.0}_{-0.6}$ times the expected SM yield. Converting this to a limit on the cross section we find at 95\% CL:
\begin{equation*}
\sigma (\Pp\Pp \rightarrow\Pp^{(*)}\PWp\PWm\Pp^{(*)} \rightarrow\Pp^{(*)}\mu^{\pm}\Pe^{\mp}\Pp^{(*)}) < 10.6\unit{fb}.
\end{equation*}
The SM prediction is $4.0\pm0.7\unit{fb}$, including the uncertainty in the contribution of proton dissociation.
The dilepton invariant mass, acoplanarity, and missing transverse energy in the two selected events are consistent with the expectation for the
sum of backgrounds and SM $\gamma\gamma\rightarrow \PWp\PWm$ signal (Fig.~\ref{fig:memudphipt0trackspt30}).

The $\pt(\mu^{\pm}\Pe^{\mp})$ distribution for events with zero extra tracks, and the extra tracks multiplicity for events with $\pt(\mu^{\pm}\Pe^{\mp})>100$\GeV,
are shown in Fig.~\ref{fig:singlemuept0trackspt30}. In the anomalous quartic gauge coupling search region $\pt(\mu^{\pm}\Pe^{\mp})>100\GeV$, zero events
are observed in data, which is consistent with the SM expectation of 0.14, dominated by $\Pp\Pp\rightarrow\Pp^{(*)}\PWp\PWm\Pp^{(*)}$.

\begin{figure}[h!t]
\centering
\includegraphics[width=6.5cm]{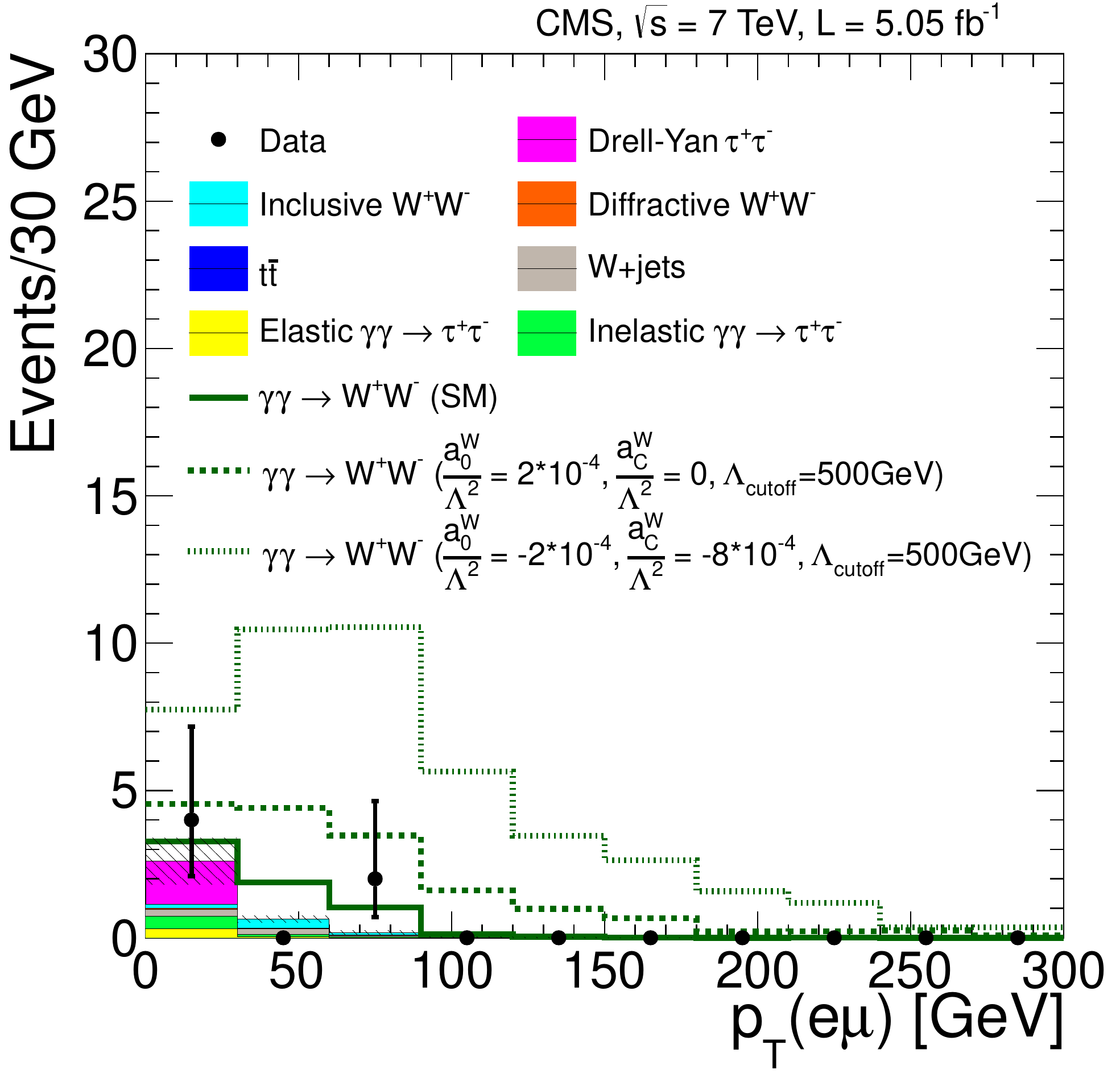}
\includegraphics[width=6.5cm]{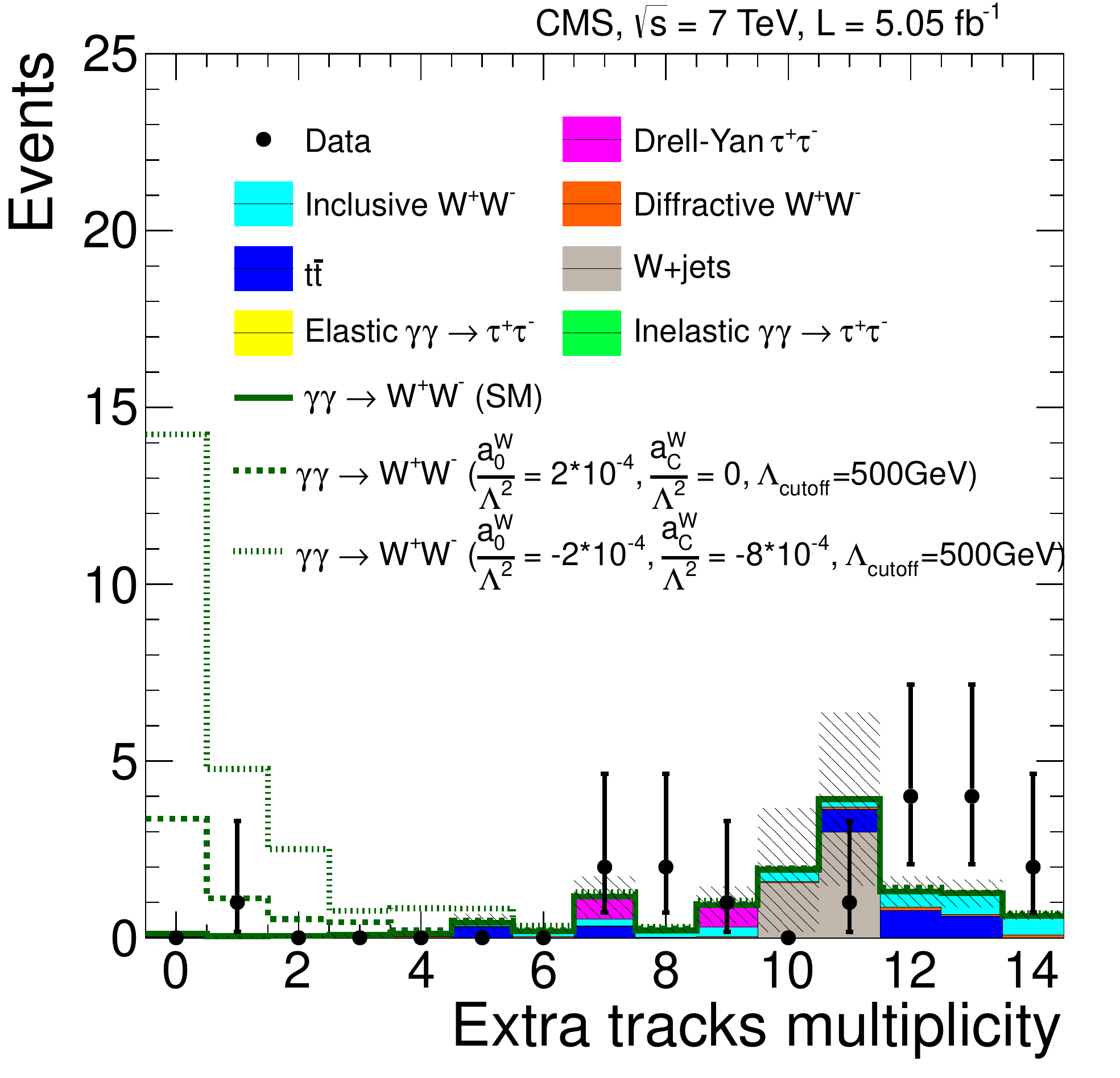}
\caption{\small{The $\pt(\mu^{\pm}\Pe^{\mp})$ distribution for events with zero extra tracks (left) and multiplicity of extra tracks for events
with $\pt(\mu^{\pm}\Pe^{\mp})>$~100\GeV (right). The backgrounds (solid histograms) are stacked with statistical uncertainties indicated by the
shaded region, the signal (open histogram) is stacked on top of the backgrounds. The expected signal is shown for the
SM $\gamma\gamma\rightarrow \PWp\PWm$ signal (solid lines) and for two representative values of the anomalous
couplings $a^{\PW}_{0}/\Lambda^{2}$ and $a^{\PW}_{C}/\Lambda^{2}$ (dotted and dashed lines).}}
\label{fig:singlemuept0trackspt30}
\end{figure}

We find that the selection efficiency does not vary strongly between the simulated SM and anomalous quartic gauge coupling samples within the detector acceptance
(Table~\ref{tab:aqgcefficiencyinacceptance}) and, therefore, set an upper limit on the partial cross section times branching fraction for
$\gamma\gamma\to \PWp\PWm \to \mu^{\pm}\Pe^{\mp}$ with $\pt(\mu,\Pe)>20\GeV$, $\abs{\eta(\mu,\Pe)}<2.4$ (for single leptons),
and $\pt(\mu^{\pm}\Pe^{\mp})>100\GeV$ for the pair. We treat the residual SM $\Pp\Pp \to\Pp^{(*)}\PWp\PWm\Pp^{(*)}$ signal as a background, resulting in
a total of $0.14\pm0.02$ expected events, and include an additional systematic uncertainty of 10\% based on the
maximum relative variation of the efficiency between the SM simulation and the samples generated with two values of the anomalous couplings.

Using the Feldman--Cousins method~\cite{Feldman:1997qc}, the 95\% CL confidence interval for the Poisson mean for signal events is [0,3.0]
if the uncertainty on the background mean is neglected. Inserting the uncertainty on the background into the frequentist
interval construction reduces the upper endpoint, as it changes the nature of the problem from a purely discrete observation with
typical over-coverage to a continuous problem with exact coverage. To avoid an effect such as this, Cousins and Highland~\cite{Cousins:1991qz}
advocated a Bayesian treatment of the nuisance parameter, which in a case such as the present one leaves the upper endpoint essentially unchanged.
This results in an upper limit on the partial cross section times branching fraction at 95\% CL with the selections
$\pt(\mu,\Pe)>20\GeV$, $\abs{\eta(\mu,\Pe)}<2.4$, and $\pt(\mu^{\pm}\Pe^{\mp})>100\GeV$:
\begin{equation*}
\sigma (\Pp\Pp \rightarrow\Pp^{(*)}\PWp\PWm\Pp^{(*)} \rightarrow\Pp^{(*)}\mu^{\pm}\Pe^{\mp}\Pp^{(*)}) < 1.9\unit{fb}.
\end{equation*}

We further investigate the behavior of the limit in different statistical approaches, with and without the systematic uncertainties included as nuisance
parameters. The limits derived from a profile likelihood method, a Bayesian method with a flat prior, the Feldman-Cousins method, the Cousins and Highland method,
and the CL$_\mathrm{S}$ method~\cite{Read:2002hq} range from 1.9 to 3.3 events at 95\% CL.

\begin{table}
\topcaption{\label{tab:aqgcefficiencyinacceptance}Signal efficiency of all trigger, reconstruction, and analysis selections, relative to the acceptance
[$\pt(\mu,\Pe)>20\GeV$, $\abs{\eta(\mu,\Pe)}<2.4$, $\pt(\mu^{\pm}\Pe^{\mp})>100\GeV$] for the SM and for four
representative values of the anomalous couplings $a^{\PW}_{0}/\Lambda^{2}$ and $a^{\PW}_{C}/\Lambda^{2}$, with and without form factors.}
\centering
\begin{tabular}{crccccc}
\hline
$a^{\PW}_{0}/\Lambda^{2}$ & [\GeVns{}$^{-2}$] & 0 & $2\times 10^{-4}$ & $-2\times 10^{-4}$  & $7.5\times 10^{-6}$ & $ 0 $                \\
$a^{\PW}_{C}/\Lambda^{2}$ & [\GeVns{}$^{-2}$] & 0 & 0                 & $-8\times 10^{-4}$  & $ 0 $                & $2.5\times 10^{-5}$ \\
$\Lambda$               & [\GeVns{}]        & -- & $500$             & $500$               & \small{No form factor}     & \small{No form factor}\\
\hline
\multicolumn{2}{l}{Efficiency}       & $30.5 \pm 5.0\%$ & $29.8 \pm 2.1\%$  & $31.3 \pm 1.8\%$ & $36.0 \pm 1.7\%$ & $36.3 \pm 1.8\%$ \\
\hline
\end{tabular}
\end{table}

The expected number of events observed as a function of the anomalous quartic gauge coupling parameters is interpolated from simulated samples
and used to construct $95\%$ CL intervals according to the Feldman--Cousins prescription. With a dipole form factor of
$\Lambda_{\text{cutoff}}=500\GeV$, the limits obtained on each anomalous quartic gauge coupling parameter with the other fixed to zero are:
\begin{align*}
-0.00015 < a^{\PW}_{0}/\Lambda^{2} < 0.00015\GeV^{-2}\,\, (a^{\PW}_{C}/\Lambda^{2} = 0, \Lambda_{\text{cutoff}}=500\GeV),\\
-0.0005 < a^{\PW}_{C}/\Lambda^{2} < 0.0005\GeV^{-2}\,\, (a^{\PW}_{0}/\Lambda^{2} = 0, \Lambda_{\text{cutoff}}=500\GeV).
\end{align*}

These limits are approximately 20 times more stringent than the best limits obtained at the Tevatron~\cite{Abazov:2013opa} with
a dipole form factor of $\Lambda_{\text{cutoff}}=500\GeV$, and approximately two orders of magnitude more stringent than the best limits obtained at
LEP~\cite{Abbiendi:2004bf,Achard:2002iz,Abdallah:2003xn}.

We perform a similar procedure to derive two dimensional limits on the $a^{\PW}_{0}/\Lambda^{2}$ and $a^{\PW}_{C}/\Lambda^{2}$ parameters. A large
number of samples generated with a fast simulation of the CMS detector~\cite{Rahmat:2012fs} is used to confirm that the signal efficiency of
all trigger, reconstruction, and analysis selections, relative to the acceptance, is flat across the anomalous quartic gauge coupling sample space,
and to derive a parameterized dependence of the cross section on the anomalous couplings. The resulting two-dimensional 95\% confidence region
is shown in Fig.~\ref{fig:2Dlimit}, including the form factor with $\Lambda_{\text{cutoff}}=500\GeV$.

We also obtain the corresponding limits without form factors. In this case the cross section is dominated by the
region of high energy $\gamma\gamma$ interactions, above the unitarity bound. This leads to one dimensional limits on each of
the anomalous couplings, with the other fixed to zero, that are much smaller than in the scenario with form factors:
\begin{align*}
-4.0  \times 10^{-6} < a^{\PW}_{0}/\Lambda^{2} < 4.0 \times 10^{-6}\GeV^{-2}\,\, (a^{\PW}_{C}/\Lambda^{2} = 0, \text{no form factor}),\\
-1.5 \times 10^{-5} < a^{\PW}_{C}/\Lambda^{2} < 1.5 \times 10^{-5}\GeV^{-2} \,\, (a^{\PW}_{0}/\Lambda^{2} = 0, \text{no form factor}).
\end{align*}

These limits are approximately two orders of magnitude more restrictive than limits obtained at the Tevatron without form factors~\cite{Abazov:2013opa}.

\begin{figure}[t!]
\centering
\includegraphics[width=8.5cm]{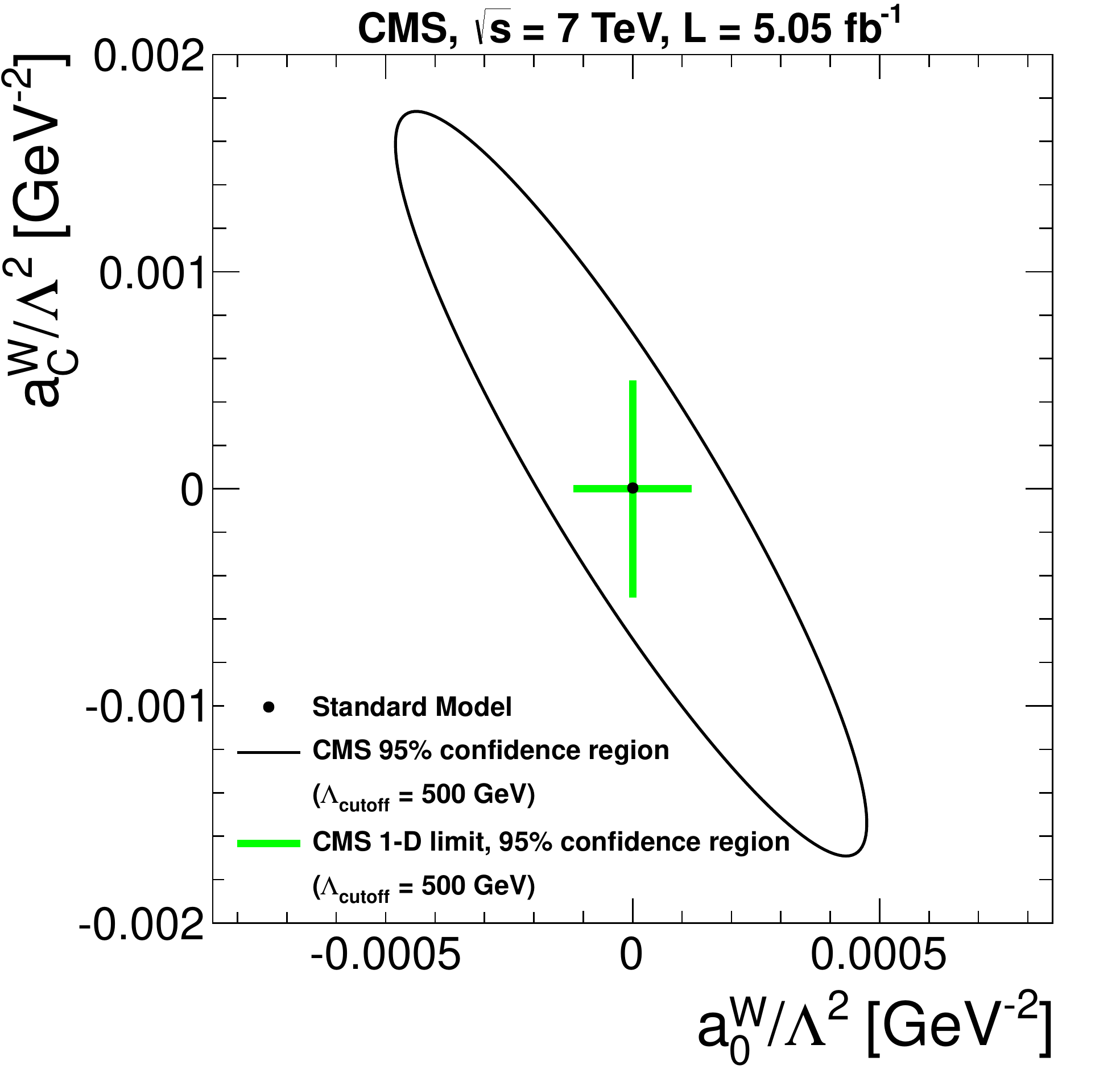}
\caption{\small{Excluded values of the anomalous coupling parameters $a^{\PW}_{0}/\Lambda^{2}$ and $a^{\PW}_{C}/\Lambda^{2}$ with $\Lambda_{\text{cutoff}}=500\GeV$.
The area outside the solid contour is excluded by this measurement at 95\%~CL, obtained for $\pt(\mu,\Pe)>20\GeV$, $\abs{\eta(\mu,\Pe)}<2.4$,
$\pt(\mu^{\pm}\Pe^{\mp})>100\GeV$. The predicted cross sections are rescaled to include the
contribution from proton dissociation.}}
\label{fig:2Dlimit}
\end{figure}

\section{Summary}
\label{summary}

A search for exclusive and quasi-exclusive two-photon production of $\PWp\PWm$ in the $\mu^{\pm}\Pe^{\mp}$ channel,
$\Pp\Pp \rightarrow\Pp^{(*)}\PWp\PWm\Pp^{(*)} \rightarrow\Pp^{(*)}\mu^{\pm}\Pe^{\mp}\Pp^{(*)}$, has been performed using 5.05\fbinv
of data collected at a center-of-mass energy of 7\TeV by the CMS detector in 2011. The efficiencies and theoretical predictions
for the signal have been checked using $\gamma\gamma\rightarrow \Pgmp\Pgmm$ events, while the backgrounds are constrained from the data
using control regions defined by the number of extra tracks and by the $\pt$ of the $\mu^{\pm}\Pe^{\mp}$ pair.

In a region sensitive to SM $\gamma\gamma \rightarrow \PWp\PWm$ production with $\pt(\mu^{\pm}\Pe^{\mp})>30\GeV$, two~events are observed,
with a background expectation of $0.84\pm0.15$. The signal expectation is $2.2\pm0.4$~events, with the uncertainty on the theory
reflecting the uncertainty on the proton dissociation contribution. The significance of the signal is $\sim$1$\sigma$, with a 95\%~CL upper limit
on the SM cross section of 10.6\unit{fb}.

In the region with $\pt(\mu^{\pm}\Pe^{\mp})>100\GeV$, where the SM contribution is expected to be small, no events
are observed. A limit is set on the partial cross section times branching fraction within the acceptance of $\pt(\mu,\Pe)>20\GeV$,
$\abs{\eta(\mu,\Pe)}<2.4$, $\pt(\mu^{\pm}\Pe^{\mp})>100\GeV$ at 95\%~CL:
\begin{equation*}
\sigma (\Pp\Pp \rightarrow\Pp^{(*)}\PWp\PWm\Pp^{(*)} \rightarrow\Pp^{(*)}\mu^{\pm}\Pe^{\mp}\Pp^{(*)}) < 1.9\unit{fb}.
\end{equation*}

We use this subsample to set limits on the anomalous quartic gauge coupling parameters, which results in values of the order of $1.5\times10^{-4}\GeV^{-2}$ for
$a^{\PW}_{0}/\Lambda^{2}$ and $5\times10^{-4}\GeV^{-2}$ for $a^{\PW}_{C}/\Lambda^{2}$, assuming a dipole form factor with the energy cutoff scale at
$\Lambda_{\text{cutoff}}=500\GeV$. These limits are approximately 20 times more stringent than the best limits obtained at the Tevatron, and approximately two
orders of magnitude more stringent than the best limits obtained at LEP. With no form factors, the limits on the anomalous quartic gauge coupling parameters would be
of order $10^{-5}\GeV^{-2}$ and below, driven by high-energy $\gamma\gamma$ interactions beyond the unitarity bound.

\section*{Acknowledgments}
\hyphenation{Bundes-ministerium Forschungs-gemeinschaft Forschungs-zentren} We congratulate our colleagues in the CERN accelerator departments for the excellent performance of the LHC and thank the technical and administrative staffs at CERN and at other CMS institutes for their contributions to the success of the CMS effort. In addition, we gratefully acknowledge the computing centres and personnel of the Worldwide LHC Computing Grid for delivering so effectively the computing infrastructure essential to our analyses. Finally, we acknowledge the enduring support for the construction and operation of the LHC and the CMS detector provided by the following funding agencies: the Austrian Federal Ministry of Science and Research and the Austrian Science Fund; the Belgian Fonds de la Recherche Scientifique, and Fonds voor Wetenschappelijk Onderzoek; the Brazilian Funding Agencies (CNPq, CAPES, FAPERJ, and FAPESP); the Bulgarian Ministry of Education, Youth and Science; CERN; the Chinese Academy of Sciences, Ministry of Science and Technology, and National Natural Science Foundation of China; the Colombian Funding Agency (COLCIENCIAS); the Croatian Ministry of Science, Education and Sport; the Research Promotion Foundation, Cyprus; the Ministry of Education and Research, Recurrent financing contract SF0690030s09 and European Regional Development Fund, Estonia; the Academy of Finland, Finnish Ministry of Education and Culture, and Helsinki Institute of Physics; the Institut National de Physique Nucl\'eaire et de Physique des Particules~/~CNRS, and Commissariat \`a l'\'Energie Atomique et aux \'Energies Alternatives~/~CEA, France; the Bundesministerium f\"ur Bildung und Forschung, Deutsche Forschungsgemeinschaft, and Helmholtz-Gemeinschaft Deutscher Forschungszentren, Germany; the General Secretariat for Research and Technology, Greece; the National Scientific Research Foundation, and National Office for Research and Technology, Hungary; the Department of Atomic Energy and the Department of Science and Technology, India; the Institute for Studies in Theoretical Physics and Mathematics, Iran; the Science Foundation, Ireland; the Istituto Nazionale di Fisica Nucleare, Italy; the Korean Ministry of Education, Science and Technology and the World Class University program of NRF, Republic of Korea; the Lithuanian Academy of Sciences; the Mexican Funding Agencies (CINVESTAV, CONACYT, SEP, and UASLP-FAI); the Ministry of Science and Innovation, New Zealand; the Pakistan Atomic Energy Commission; the Ministry of Science and Higher Education and the National Science Centre, Poland; the Funda\c{c}\~ao para a Ci\^encia e a Tecnologia, Portugal; JINR (Armenia, Belarus, Georgia, Ukraine, Uzbekistan); the Ministry of Education and Science of the Russian Federation, the Federal Agency of Atomic Energy of the Russian Federation, Russian Academy of Sciences, and the Russian Foundation for Basic Research; the Ministry of Science and Technological Development of Serbia; the Secretar\'{\i}a de Estado de Investigaci\'on, Desarrollo e Innovaci\'on and Programa Consolider-Ingenio 2010, Spain; the Swiss Funding Agencies (ETH Board, ETH Zurich, PSI, SNF, UniZH, Canton Zurich, and SER); the National Science Council, Taipei; the Thailand Center of Excellence in Physics, the Institute for the Promotion of Teaching Science and Technology of Thailand and the National Science and Technology Development Agency of Thailand; the Scientific and Technical Research Council of Turkey, and Turkish Atomic Energy Authority; the Science and Technology Facilities Council, UK; the US Department of Energy, and the US National Science Foundation.

Individuals have received support from the Marie-Curie programme and the European Research Council and EPLANET (European Union); the Leventis Foundation; the A. P. Sloan Foundation; the Alexander von Humboldt Foundation; the Belgian Federal Science Policy Office; the Fonds pour la Formation \`a la Recherche dans l'Industrie et dans l'Agriculture (FRIA-Belgium); the Agentschap voor Innovatie door Wetenschap en Technologie (IWT-Belgium); the Ministry of Education, Youth and Sports (MEYS) of Czech Republic; the Council of Science and Industrial Research, India; the Compagnia di San Paolo (Torino); the HOMING PLUS programme of Foundation for Polish Science, cofinanced by EU, Regional Development Fund; and the Thalis and Aristeia programmes cofinanced by EU-ESF and the Greek NSRF.

\bibliography{auto_generated}   % will be created by the tdr script.

\cleardoublepage \appendix\section{The CMS Collaboration \label{app:collab}}\begin{sloppypar}\hyphenpenalty=5000\widowpenalty=500\clubpenalty=5000\textbf{Yerevan Physics Institute,  Yerevan,  Armenia}\\*[0pt]
S.~Chatrchyan, V.~Khachatryan, A.M.~Sirunyan, A.~Tumasyan
\vskip\cmsinstskip
\textbf{Institut f\"{u}r Hochenergiephysik der OeAW,  Wien,  Austria}\\*[0pt]
W.~Adam, T.~Bergauer, M.~Dragicevic, J.~Er\"{o}, C.~Fabjan\cmsAuthorMark{1}, M.~Friedl, R.~Fr\"{u}hwirth\cmsAuthorMark{1}, V.M.~Ghete, N.~H\"{o}rmann, J.~Hrubec, M.~Jeitler\cmsAuthorMark{1}, W.~Kiesenhofer, V.~Kn\"{u}nz, M.~Krammer\cmsAuthorMark{1}, I.~Kr\"{a}tschmer, D.~Liko, I.~Mikulec, D.~Rabady\cmsAuthorMark{2}, B.~Rahbaran, C.~Rohringer, H.~Rohringer, R.~Sch\"{o}fbeck, J.~Strauss, A.~Taurok, W.~Treberer-Treberspurg, W.~Waltenberger, C.-E.~Wulz\cmsAuthorMark{1}
% ...
\vskip\cmsinstskip
% ...
1:~~Also at Vienna University of Technology, Vienna, Austria\\
2:~~Also at CERN, European Organization for Nuclear Research, Geneva, Switzerland\\
% ...

\end{sloppypar}
\end{document}